\documentclass[12pt, draftcls, onecolumn]{IEEEtran}

\newcommand{\DQ}[1]{\textcolor{blue}{{#1}}}

\newcommand{\M}[1]{\textcolor{teal}{{#1}}}

\def\x{\bm{x}}\def\w{\bm{w}}\def\0{\bm{0}}\def\bu{\bm{u}}

\usepackage[latin1]{inputenc}																							%
\usepackage{type1cm}																										%
\usepackage{type1ec}																										%
\usepackage[T1]{fontenc}																								%
\usepackage{dsfont}																										%
\usepackage{booktabs}																									%
\usepackage{array}																											%
\usepackage{enumerate}																									%
\usepackage[table]{xcolor}																								%
\usepackage{makeidx}																										%
\usepackage{fancyhdr}																									%
\usepackage{bm}																												%
\usepackage{lastpage}																										%
\usepackage{multicol}																										%
\usepackage{amsmath, amssymb, amsfonts}																	%
\usepackage{ifthen}																											%
\usepackage{ifpdf}																											%
\usepackage{framed}																										%
\usepackage[amsmath,thmmarks,framed,hyperref]{ntheorem}										%
\DeclareMathAlphabet{\mathpzc}{OT1}{pzc}{m}{it}		
\usepackage{amsmath,amsfonts,amssymb,booktabs}													%
\usepackage{amstext}																										%
\usepackage{psfrag} 																										%
\usepackage{float} 																											%
\usepackage{cite} 																											%
\usepackage[dvips]{graphicx} 																							%
\usepackage{subfloat} 																									%
\usepackage{stackrel} 																										%
\usepackage{graphicx} 																									%
\usepackage{textcomp} 																									%
\usepackage{color} 																											%
\usepackage{bbm} 																											%
\usepackage{lettrine}																										%
\usepackage{dsfont} 																										%
\usepackage{multirow}																									%
\usepackage[section,verbose]{placeins} 																			%
\usepackage{url}
\usepackage{tikz}																											%
\usetikzlibrary{arrows,automata,matrix,fit,positioning,calc}             								%
\usepackage{pgfplots}																										%
\usepackage[latin1]{inputenc}																							%
\usepackage{placeins}

\usepackage[numbers,sort&compress]{natbib}

\newcommand{\bdel}{\boldsymbol{\delta}}
\newcommand{\bphi}{\boldsymbol{\phi}}

\usepackage{framed}																							  %
\usepackage[amsmath,thmmarks,framed,hyperref]{ntheorem}							  %
\newcounter{generalCounter}
\theoremclass		{Theorem}
\theoremstyle		{plain}
\theoremheaderfont	{\normalfont\bfseries}
\theorembodyfont	{\normalfont}
\newshadedtheorem	{definition}		[generalCounter]	{Definition}
\newshadedtheorem	{theorem}			[generalCounter]	{Theorem}
\newshadedtheorem	{proposition}	[generalCounter]	{Proposition}
\newshadedtheorem	{assumption}	[generalCounter]	{Assumption}
\newshadedtheorem	{hypothesis}		[generalCounter]	{Hypothesis}
\newshadedtheorem	{corollary}			[generalCounter]	{Corollary}
\newshadedtheorem	{lemma}			[generalCounter]	{Lemma}
\newshadedtheorem	{remark}			[generalCounter]	{Remark}
\newshadedtheorem	{question}			[generalCounter]	{Q.}
\newshadedtheorem	{discussion}		[generalCounter]	{Discussion}
\usepackage{setspace}

\title{ On the trade-off between control performance and communication cost in event-triggered control }

\author{\IEEEauthorblockN{Burak Demirel\IEEEauthorrefmark{1}\IEEEauthorrefmark{2}, Vijay Gupta\IEEEauthorrefmark{3}, Daniel E. Quevedo\IEEEauthorrefmark{4} and Mikael Johansson\IEEEauthorrefmark{1}}
\thanks{
				\IEEEauthorblockA{\IEEEauthorrefmark{1}ACCESS Linnaeus Centre, School of Electrical Engineering, KTH Royal Institute of Technology, Osquldas vag~10, SE 10044 Stockholm, Sweden. } 
				\IEEEauthorblockA{\IEEEauthorrefmark{3}Department of Electrical Engineering, University of Notre Dame, IN 46556 USA. } 
				\IEEEauthorblockA{\IEEEauthorrefmark{4}School of Electrical Engineering \& Computer Science, Callaghan, University of Newcastle, NSW 2308, Australia. } 
				}
\thanks{
\IEEEauthorrefmark{2} Corresponding author. E-mail: burak.demirel@ee.kth.se. 
A preliminary version of parts of this work was presented at the $12^{th}$ Biannual European Control Conference, see~\cite{DGJ:13}. }
			}
\IEEEoverridecommandlockouts

\begin{document}

\maketitle




\begin{abstract}
We consider a stochastic system where the communication between the controller and the actuator is triggered by a threshold-based rule. The communication is performed across an unreliable link that stochastically erases transmitted packets. To decrease the communication burden, and as a partial protection against dropped packets, the controller sends a sequence of control commands to the actuator in each packet. These commands are stored in a buffer and applied sequentially until the next control packet arrives.  In this context, we study dead-beat control laws and compute the expected linear-quadratic loss of the closed-loop system for any given event-threshold. Furthermore, we provide analytical expressions that quantify the trade-off between the communication cost and the control performance of event-triggered control systems.
Numerical examples demonstrate the effectiveness of the proposed framework.
\end{abstract}



\begin{keywords}
Event-triggering algorithms; Linear systems; Communication networks; Packet loss; Networked control
\end{keywords}


\IEEEpeerreviewmaketitle

\section{Introduction}\label{sec:Introduction}
\lettrine{M}{any} textbooks on sampled-data control, e.g.,~\cite{AsW:97, ChF:95}, exhibit periodic control as the unique way to implement feedback control laws on digital platforms. The rationale behind this choice is that there exists a well-developed theory that allows to analyze the stability of closed-loop systems, to evaluate their control performance, and to design optimal controllers. Despite all these advantages, periodic implementations might lead to an inefficient use of the communication medium. For instance, transmitting the same actuator value repeatedly when the system is at rest at the desired state is undoubtedly a waste of communication resources. In contrast, event-triggered implementations of feedback control laws adapt the use of the communication channel to the needs of the physical system. Since event-triggered implementations of control laws are often able to achieve a satisfactory performance using significantly reduced communication rates (see e.g., the tutorial paper~\cite{HJT:12} and the references therein) they have emerged as an attractive alternative approach to the traditional periodic implementations. A reduced communication rate decreases the energy consumption at the transmitter side and reduces the network congestion when the communication takes place over a shared medium. For all these reasons, event-triggered implementations have been receiving an increasing attention in many applications including, e.g., control over communication networks~\cite{HvK:02, RSJ:11, RSJ:12, RSJ:13, AMA+:14}, multi-agent systems~\cite{DFJ:12}, distributed optimization~\cite{Lem:10}, and embedded control systems~\cite{Tab:07}.

\subsection{Related Work}

A number of different event-triggering mechanisms have been proposed in the literature. These can be broadly categorized as Lyapunov-based~\cite{Tab:07,HDT:13,AnT:10,Lem:10,VMB:09}, model-based~\cite{LuL:10,HeD:13,MoA:04}, or threshold-based~\cite{OMT:02,HSB:08,XGA:12}. It is widely recognized that event-triggered implementations can decrease the communication load in a networked control system compared to periodic ones while still guaranteeing closed-loop stability and performance~\cite{AsB:99,AsB:02,HJC:08}. However, quantifying the expected transmission rate of such implementations for specified performance in closed-loop is challenging.
A notable work in quantifying such a relation is that of \AA str\"{o}m and Bernhardsson~\cite{AsB:02} who focused on the threshold-based event-triggered implementation of an impulse control of a single integrator system under Wiener process disturbances. They established that the event-based implementation gives a better performance than the traditional periodic implementation in terms of the state variance. Similarly, Henningsson et al.~\cite{HJC:08} proposed event-triggered control scheme and compared the achievable performance by using this scheme with respect to periodic control. Rabi~\cite{Rab:06} designed the joint optimal event-triggering mechanism and the control law to minimize the average energy of the state signal. However, the authors of~\cite{AsB:02,HJC:08,Rab:06} consider a scalar system to be controlled, which affords significant simplifications of the analysis. More recently, Meng and Chen~\cite{MeC:12} extended the work of~\cite{AsB:02} to a class of second-order stochastic systems, and showed that, for the same average transmission rate, the event-based impulse control outperforms the periodic one. Nevertheless, as stated in~\cite{MeC:12}, extending the proposed technique to consider more general system dynamics remains an open and challenging problem.

The analysis becomes even more involved when the network is unreliable. 
Blind and Allg\"{o}wer~\cite{BlA:12} extended the work of~\cite{AsB:02} to the case where transmissions from the sensor to the controller take place over an unreliable link. They analytically derived the control cost and the expected inter-event times for different packet loss rates. Rabi and Johansson~\cite{RaJ:09} designed the optimal impulse control and the level triggering mechanism under packet losses with multiple loops sharing a common network. From a different perspective, Quevedo et al.~\cite{QGM+:14} analyzed the stability of an event-triggered implementation of a controller in the presence of packet losses and limited processing resources. However, in all these works, the unreliable channel appears between sensor and controller, while the controller-actuator communication is lossless.

In this paper, we study a threshold-based event-triggered control loop with unreliable communication between the controller and actuator. We develop a framework for quantifying the closed-loop performance and the expected communication rate between the controller and actuator without any restriction on the order of the system under control. As a control strategy, we apply packetized control, which is a well-known technique in the presence of unreliable communication channels; see, e.g.,~\cite{QuN:11, QuN:12, MaG:12}. There are only a few studies in the literature on the blend of event-triggered control and packetized control; e.g.,~\cite{ZLR:10, DFH:13,QGM+:14}. To the best of our knowledge, this paper is the first attempt to characterize analytically the trade-off between the expected communication rate and the control performance in the event-triggered control using a packetized control strategy.

\subsection{Main contributions of the paper}

The main contribution of this paper is to provide analytical expressions that quantify the trade-off between communication cost and control performance in packetized event-triggered control systems with unreliable controller-actuator communication. We consider a linear stochastic system where the communication between the controller and the actuator is dictated by a threshold-based event-triggering algorithm. The communication is across an analog erasure channel that stochastically erases transmitted data at any time step. We consider the cheap control case with the controller chosen to minimize the state variance. Using a Markov renewal process-based framework, we are able to establish analytical expressions for the expected communication rate and the control performance as measured by a linear-quadratic cost.

\subsection{Outline of the paper}
The rest of the paper is organized as follows. In Section~\ref{sec:ProblemFormulation}, we introduce models of the process and communication channel, describe the control architecture and the event-triggering rule and specify measures of communication cost and control performance. Sections~\ref{sec:MainResults} and~\ref{sec:MainResultsHigherOrder} presents our main results, first for scalar, and then for higher-order systems. Numerical examples in Section~\ref{sec:NumExamples} illustrate the power of our framework. Finally, concluding discussions and directions for future work are given in Section~\ref{sec:Conclusion}. The appendix provides detailed proofs of the main results.

\subsection{Notation}
We write $\mathbb{N}$ for the positive integers and $\mathbb{N}_{0}$ for $\mathbb{N}\cup\{0\}$, and we use $\mathbb{R}$ for the real numbers. Let $\mathbb{R}_{\succeq 0}^{n}$ denote the set of non-negative real vectors of dimension $n$, and $\mathbb{R}^{n}$ denote the set of real vectors of dimension $n$. Vectors are written in bold lower case letters and matrices in capital letters. If $\bm{u}$ and $\bm{v}$ are two vectors in $\mathbb{R}^{n}$, the notation $\bm{u}\leq\bm{v}$ corresponds to the component-wise inequality. The set of all real symmetric positive semi-definite matrices of dimension $n$ is denoted by $\mathbb{S}_{\succeq 0}^{n}$. We let $\mathbf{0}_{n}$ be the $n$--dimensional column vectors of all zeros, $\mathbf{1}_{n}$ be the vectors of all ones. For any given $\bm{x}\in\mathbb{R}^{n}$, the $\ell_{\infty}$ norm is defined by $\parallel\bm{x}\parallel_{\infty}=\max\limits_{1\leq i\leq n}\vert x_{i} \vert$. For a square matrix $A$, $\mbox{Tr}(A)$ denotes its trace, $\vert A\vert$ its determinant and $\lambda_{\max}(A)$ its maximum eigenvalue in terms of magnitude. In symmetric block matrices, we use $\star$ to represent elements implied by symmetry. Let $X = \mathtt{lyap}( A,Q )$ denote the positive semi-definite solution of the discrete Lyapunov matrix equation: $AXA^{\intercal} - X + Q = 0$, for any given $Q\in\mathbb{S}_{\succeq 0}^{n}$ and $A\in\mathbb{R}^{n\times n}$ with $\lambda_{\max}(A)< 1$.
The notation$\{x_{k}\}_{k\in\mathcal{K}}$ stands for $\{x(k) : k\in\mathcal{K}\}$, where $\mathcal{K}\subseteq\mathbb{N}_{0}$. We use the symbol $\mathbbm{1}_{\{x_{k}\in A\}}$ to denote the indicator function of the set $A$. The probability of an event $\Omega$ is denoted by $\mathbf{P}\big[\Omega\big]$ and the conditional probability of $\Omega$ given $\Gamma$ is written as $\mathbf{P}\big[\Omega\mid\Gamma\big]$. When $\chi$ is a stochastic variable, $\mathbf{E}[ \chi ]$ stands for the expectation of $\chi$, $\mathbf{Var}[ \chi ]$ stands for the variance of $\chi$ and $\mathbf{Cov}[ \chi ]$ stands for the covariance of $\chi$.
An $n$--dimensional vector of real-valued random variables $\x=[ x_{1}~\cdots~x_{n} ]^{\intercal}$ follows a multivariate normal distribution with mean $\boldsymbol{\mu}\in\mathbb{R}^{n}$ and covariance matrix $\Sigma\in\mathbb{S}_{\succeq 0}^{n}$, denoted by $\mathcal{N}\big( \boldsymbol{\mu}, \Sigma \big)$, if its probability density function is given by
\begin{equation}
	f(\x;\boldsymbol{\mu}, \Sigma) = \frac{1}{(2\pi)^{\frac{n}{2}}\vert\Sigma\vert^{\frac{1}{2}}} e^{-\frac{1}{2}(\x-\boldsymbol{\mu})^{\intercal}\Sigma^{-1}(\x-\boldsymbol{\mu})} \;.
\end{equation}
The cumulative distribution function $F(\boldsymbol{\epsilon} ;\boldsymbol{\mu},\Sigma)$ is defined as 
\begin{align}
	F(\boldsymbol{\epsilon} ;\boldsymbol{\mu},\Sigma) \triangleq 
	\frac{1}{(2\pi)^{\frac{n}{2}}\vert\Sigma\vert^{\frac{1}{2}}}\int_{\boldsymbol{\epsilon}}^{\infty}e^{-\frac{1}{2}(\x-\boldsymbol{\mu})^{\intercal}\Sigma^{-1}(\x-\boldsymbol{\mu})}d\x \;.
\label{eq:GaussianIntegral}
\end{align}
Suppose one or more variates of multivariate normal random variable $\x$ are subject to one-sided or two-sided truncation, i.e., $\boldsymbol{\epsilon}^{-}\leq\x\leq\boldsymbol{\epsilon}^{+}$. Then, $\x$ has a truncated normal distribution and its probability density function is given by
\begin{equation}
	f(\x;\boldsymbol{\mu}, \Sigma, \boldsymbol{\epsilon}^{-}, \boldsymbol{\epsilon}^{+}) = \frac{e^{-\frac{1}{2}(\x-\boldsymbol{\mu})^{\intercal}\Sigma^{-1}(\x-\boldsymbol{\mu})}}{\int_{\boldsymbol{\epsilon}^{-}}^{\boldsymbol{\epsilon}^{+}}e^{-\frac{1}{2}(\x-\boldsymbol{\mu})^{\intercal}\Sigma^{-1}(\x-\boldsymbol{\mu})}d\x} \;.
\end{equation}
There are many techniques for sampling from truncated multivariate normal distributions, such as techniques based on the accept-reject algorithm and Gibbs sampling~\cite{Rob:95}. The {\tt R} package {\tt mtmvnorm}~\cite{MaW:09} provides several efficient methods to work with truncated random variables.


\section{Problem formulation} \label{sec:ProblemFormulation}
This section summarizes the control architecture of our event-triggered control scheme and introduces the assumptions under which we will  develop the performance analysis.

\subsection{Control architecture}

\begin{figure}\centering
	\includegraphics[scale=0.17]{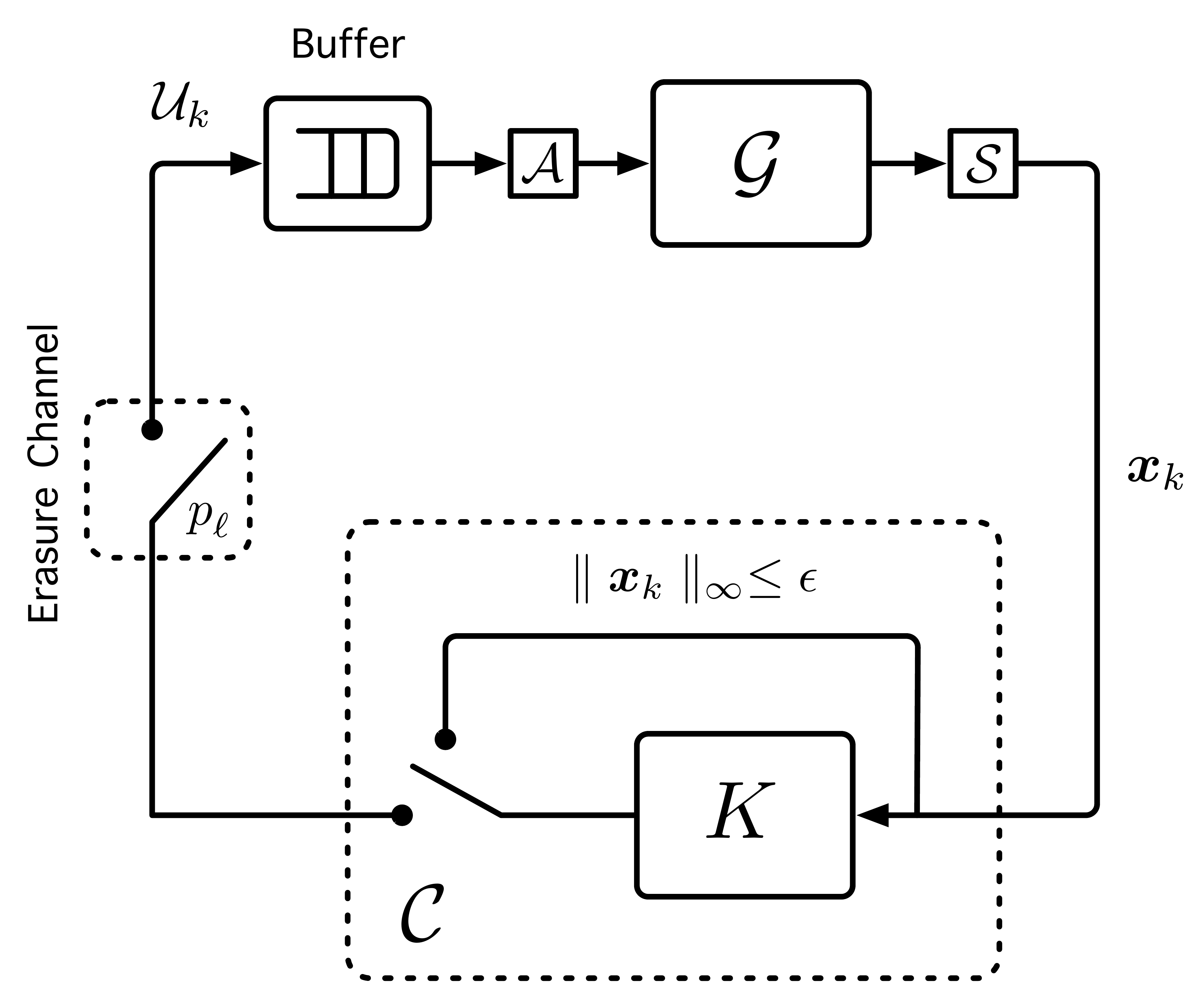}
	\caption{Block diagram of event-triggered control system with a (linear) plant $\mathcal{G}$, a controller $\mathcal{C}$, a sensor $\mathcal{S}$, an actuator $\mathcal{A}$, a buffer with queue size $\nu-1$, a comparator with event-triggering rule $\parallel x_{k} \parallel_{\infty} >\epsilon$, and an unreliable communication link.} 
	\label{fig:Block_Diagram_ETC_VectorCase}
\end{figure}

We consider the feedback control loop shown in Fig.~\ref{fig:Block_Diagram_ETC_VectorCase}. A physical plant $\mathcal{G}$, whose dynamics can be represented by a linear stochastic system, is being controlled. A sensor $\mathcal{S}$ takes periodic samples of the plant state $\x_{k}$ and transmits these to the controller node. The controller $\mathcal{C}$ is event-triggered and computes new actuation commands only at times when the plant state satisfies the event-triggering condition $\parallel \x_{k} \parallel_{\infty}>\epsilon$ for a given threshold $\epsilon> 0$. The communication between the controller $\mathcal{C}$ and the actuator $\mathcal{A}$ is lossy, and control packets are dropped at any time step independently of each other, with probability $p_{\ell}\in(0,1)$. As partial protection against these losses, the controller sends a sequence of predicted commands in each packet. The predicted commands are placed in a buffer at the actuator. In the absence of new control packets, the actuator reads the predicted control command for the current time from the buffer and applies this input to the plant. In this context, we are interested in deriving analytical performance guarantees, both in terms of control performance and the number of transmission attempts on the communication link between controller and actuator.
%
%

\subsection{Process model}\label{sec:ProcessModel_HOS}

The dynamics of the plant $\mathcal{G}$ can be described by the stochastic discrete-time linear system:
\begin{align}
	\x_{k+1} = A\x_{k} + B\bu_{k} + \w_{k},
	\label{eqn:SystemModel_VectorCase}
\end{align}
where $\x_{k}\in\mathbb{R}^{n}$ is the state variable, $\bu_{k}\in\mathbb{R}^{m}$ is the control signal, and $\w_{k}\in\mathbb{R}^{n}$ is a discrete-time zero-mean Gaussian white noise with covariance $\Sigma_{w}\in\mathbb{S}^{n}_{\succeq 0}$, i.e., $\w_{k}\sim\mathcal{N}(\mathbf{0}_{n},\Sigma_{w})$. The initial state $\x_{0}$ is modeled as a random variable having a normal distribution with zero mean and covariance $\Sigma_{0}\in\mathbb{S}^{n}_{\succeq 0}$, i.e., $\x_{0}\sim\mathcal{N}(\mathbf{0}_{n},\Sigma_{0})$. The process noise $\{\w_{k}\}_{k\in\mathbb{N}_{0}}$ is independent of the initial condition $\x_{0}$.

The system matrix $A\in\mathbb{R}^{n\times n}$ and the input matrix $B\in\mathbb{R}^{n\times m}$ are constant, and $B$ is assumed to be of full column-rank. Furthermore, the system~\eqref{eqn:SystemModel_VectorCase} with $\w_{k}=\mathbf{0}_{n}$ for all $k\in\mathbb{N}_{0}$ is assumed to be completely $\nu$-step controllable for some $\nu\leq n$. In other words, for every $\x_{k}\in\mathbb{R}^{n}$, there exists a control sequence $\mathcal{U}=\{ \bu_{k}, \bu_{k+1}, \cdots, \bu_{k+\nu-1} \}$ that transfers the state from $\x_{k}$ to the origin in $\nu$ time steps. When $A$ is non-singular (which it will be if it is obtained by sampling a delay-free continuous-time linear system), the system~\eqref{eqn:SystemModel_VectorCase} is completely $\nu$-step controllable~\cite{Ore:81} if and only if
	\begin{align*}	
		\mbox{rank}\big[ A^{-1}B, A^{-2}B, \cdots, A^{-\nu}B \big] = n\;.
	\end{align*}

\subsection{Controller design and performance criterion}\label{sec:ControllerDesign}

We quantify the closed-loop performance in terms of the quadratic cost
\begin{align}
	J_{\infty} = \lim_{N\rightarrow\infty}\frac{1}{N}\sum_{k=0}^{N-1}\Big( \x_{k}^{\intercal}Q_{x}\x_{k}^{} + \rho \bu_{k}^{\intercal}Q_{u}\bu_{k}^{} \Big)
	\label{eqn:ControlLossFunction}
\end{align}
for given symmetric positive semidefinite matrix $Q_{x}$ and symmetric positive definite matrix $Q_{u}$. We are particularly interested in the case when $\rho\rightarrow 0$, sometimes called {\it the cheap control scenario}; see, e.g.,~\cite{Hes:09}. It is well known that the optimal controller for the cheap control scenario is a dead-beat controller which ensures that, in the absence of process noise, the state converges to zero in a finite number of steps~\cite{AsW:97}. 
Our analysis framework considers a standard linear dead-beat controller
\begin{align}
	\bu_{k} = K\x_{k} \;,
	\label{eqn:DeadbeatControl}
\end{align}
and a cost of the form~\eqref{eqn:ControlLossFunction}. When the system is completely $\nu$-step controllable, one can always find such a controller $K$ that drives the system state to zero in $\nu$ steps (see e.g.,~\cite{Ore:81}).

We use a packetized dead-beat controllerto reduce transmissions over the communication channel and to guard against losses. If the event-triggering rule leads to the controller executing the control algorithm at time $k\in\mathbb{N}_{0}$, it computes and transmits a sequence of control commands
\begin{align}
	\mathcal{U}_{k} = \{ K\x_{k}, K(A+BK)\x_{k}, \cdots, K(A+BK)^{\nu-1}\x_{k} \}
	\label{eqn:ControlSequence}
\end{align}
which would transfer the process state of~\eqref{eqn:SystemModel_VectorCase} from $\x_{k}$ to the origin in at most $\nu\leq n$ time steps in the absence of process noise. We assume the presence of a buffer of length $\nu-1$ at the actuator. When a new set of control actions arrives from the controller, the actuator immediately applies the first control action in the set, and stores the rest of the control actions in a buffer, see Fig~\ref{fig:Block_Diagram_ETC_VectorCase}. In the next $\nu-1$ time steps, the controller issues no transmissions even if the event-triggered condition is met. Rather, the actuator applies the control commands sequentially from the buffer. If the buffer is empty, the actuator applies zero input (cf.~\cite{Sch:09}). Note that this is consistent with the dead-beat assumption: in the absence of noise, the state would be at the origin after $\nu$ steps and~\eqref{eqn:DeadbeatControl} would evaluate to zero.

After each successful packet transmission, the controller is switched off for $\nu-1$ time steps, and then it is switched on again. The controller uses a simple threshold-based rule
\begin{align}
	\parallel \x_{k} \parallel_{\infty} > \epsilon \label{eqn:threshold_condition}
\end{align}
to determine if a new control sequence should be computed, and a transmission should be attempted between the controller and actuator. We assume that the communication primitive offers reliable acknowledgements, so that the controller node knows if a transmission attempt to the actuator node was successful. In addition, we introduce a time-out mechanism where, if the number of samples since the last successful transmission exceeds a time-out value of $T$, the controller will attempt to transmit new data to the actuator even if  the plant state does not satisfy the triggering condition~\eqref{eqn:threshold_condition}.
Such a time-out mechanism, admittedly essential to our analysis, is also important in event-triggered control system to guard against faulty components. It is worth noting that the time-out $\mathrm{T}$ is a design parameter which is not directly related to $\nu$.

\subsection{Communication channel}

The communication channel between the controller and actuator is lossy, and packets transmitted from the controller to the actuator are dropped with a probability $p_{\ell}\in(0,1)$, independent of all other events in the system. If a transmitted control packet is dropped, the transmission of a new control command will be attempted in the next sampling instant, irrespectively of the state evolution.

\subsection{Discussion}

At any time $k\in\mathbb{N}_{0}$, the system can operate in one of three modes: (a) we choose not to transmit any information, but let the actuator use data from the buffer; (b) we attempt to transmit a new packet, but it is dropped by the communication link; or (c) we succesfully transmit a new control sequence from the controller to the actuator.

At first glance, it might appear that the system~\eqref{eqn:SystemModel_VectorCase} with a packetized dead-beat controller~\eqref{eqn:ControlSequence} is a Markovian Jump Linear System (MJLS). However, this is not true since the transmission of control updates under the event-triggering rule depends on the plant state (i.e., the Markov jump process depends on the continuous state variable). Thus, traditional techniques for analyzing the long-run average of control performance, such as~\cite{CFM:05}, cannot be applied. In fact, our system fits into the class of the state-dependent random-time switched systems, which are hard to analyze in general; see e.g.,~\cite{ZYL:13}. Nonetheless, we, here, propose a technique which allows us to analyze a similar class of problems when the control sequence satisfies a dead-beat assumption.

\section{Event-triggered control of first-order systems}\label{sec:MainResults}

For pedagogical ease, we first restrict our attention to the case when the plant ${\mathcal G}$ is a first-order system. In other words, $m=n=1$. To emphasize that $A$ and $B$ are scalars, we write $A=a$, $B=b$ and the covariance matrices as $\Sigma_{w}=\sigma_{w}^{2}$ and $\Sigma_{0}=\sigma_{0}^{2}$. The corresponding dead-beat controller is $K=-a/b$ and since $\nu=1$, no buffer is needed at the actuator. 

In Section~\ref{sec:ETC_FirstOrderSys}, we consider a reliable communication channel between the controller and actuator (i.e., $p_{\ell}=0$). The extenion to packet dropouts is given in Section~\ref{sec:ETC_FirstOrderSys_w_Loss}. Proofs of technical results are presented in the Appendix.

\subsection{Control over perfect channel} \label{sec:ETC_FirstOrderSys}

\begin{figure}\centering
	\includegraphics[scale=0.35]{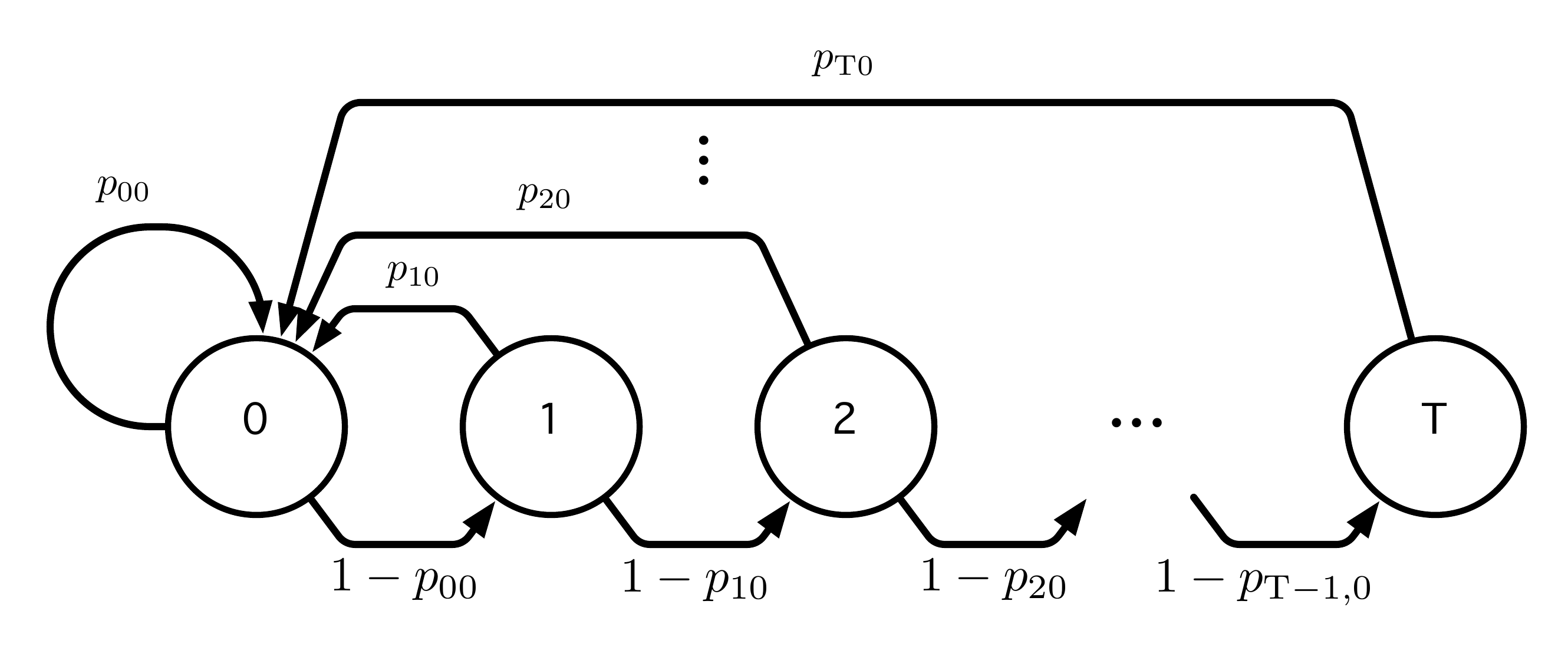}
	\caption{The transition graph of Markov chain $\{r_{k}\}$ with transition probabilities $p_{ij}$; this Markov chain has $\mathrm{T}+1$ states labelled $0,\cdots,\mathrm{T}$.} 
	\label{fig:Markov_Chain_wo_Packet_Loss}
\end{figure}

Assume that the controller-actuator communication is reliable, that is $p_{\ell}=0$ and that ${\mathcal G}$ is a first-order system. For convenience, we rewrite the evolution of the process $\{x_{k}\}$ as:
\begin{align}
	x_{k+1} = \phi_{k}(r_{k})x_{k} + w_{k} \;, \label{eq:Stochastic_System_2}
\end{align}
with initial condition $x_{0}$ \big(correspondingly $r_{0}\in\{0,1\}$\big) and
\begin{align}
	\phi_{k}(r_{k}) = 
	\begin{cases}
		0 & \text{if}~r_{k}=0 \;, \\
		a & \text{otherwise} \;,
	\end{cases} \label{eq:Stochastic_System_2_extra}
\end{align}
where $x_{k}\in\mathbb{R}$ denotes the state of the process, and $w_{k}$ is a zero-mean Gaussian white noise process with positive variance $\sigma_{w}^{2}$ and $w_{k}$ is independent of $x_{0}$.
The process $\{r_{k}\}_{k\in\mathbb{N}_{0}}$ describes how many time steps ago the latest transmission of a control packet occurred. Whenever $r_{k}=0$, a control packet is transmitted from the controller to the actuator. It is also worth noting that the number of time steps between two consecutive transmissions is bounded by the time-out interval $\mathrm{T}\in\mathbb{N}_{0}$. To this end, the evolution of the process $\{r_{k}\}$ is defined by
\begin{align}
	r_{k+1} = 
	\begin{cases}
		0 & \text{if}~\vert x_{k+1}\vert>\epsilon~\text{or}~r_{k}>\mathrm{T} \;, \\
		r_{k} + 1 & \text{otherwise} \;.
	\end{cases} \label{eq:IncStep}
\end{align}
Since the dead-beat control resets the plant state $x_{k}$ whenever $\{r_{k}=0\}$, the stochastic recurrence equation~\eqref{eq:Stochastic_System_2} can be re-written as
\begin{align}
	x_{k+1} = \sum_{l=0}^{r_{k}}a^{l}w_{k-l} \;. \label{eq:IncStep2}
\end{align}
This implies that the probability of an event-threshold crossing at any time $k$ only depends on $a$, $\sigma_0^2$, $\sigma^2_w$ and the current value of $r_k$. Hence, the evolution of the system can be represented by the Markov chain $\{r_{k}\}$ shown in Fig.~\ref{fig:Markov_Chain_wo_Packet_Loss}.

\begin{lemma} \label{lem:MarkovChain1}
The process $\{ r_{k} \}_{k\in\mathbb{N}_{0}}$ is an ergodic, time-homogeneous Markov chain (i.e. irreducible, aperiodic and positive recurrent) with a finite state space $\mathcal{B}=\{ 0,1, \cdots, \mathrm{T} \}$ and it has a unique invariant distribution $\boldsymbol{\pi}$ such that $\pi_{i}>0$ for all $i\in\mathcal{B}$.
\end{lemma}

To characterize the transition probabilities 
\begin{equation*}
	p_{ij} \triangleq \mathbf{P}(r_{k+1}=j \mid r_{k}=i) \;,
\end{equation*}
of the Markov chain $\{ r_k\}$, we introduce the random processes 
\begin{align}
	\xi_{k}(i)= \sum_{j=0}^{i}a^{j}w_{k-j}\;,
	\label{eq:RandomVarA}
\end{align}
for all $i\in\{0, 1, \cdots,\mathrm{T}-1\}$. The probability density functions of $\xi_{k}(i)$ are time-invariant since the noise $\{w_{k}\}_{k\in\mathbb{N}_{0}}$ is white and stationary. Hence, we can drop the time $k$ index to simplify notation. Observe that the vector-valued random variable $\boldsymbol{\xi}_{i}\triangleq \big[ \xi_{0}~\xi_{1}~\cdots~\xi_{i} \big]^{\intercal}$ has a multi-variate normal distribution with mean $\mathbf{E}\big[ \boldsymbol{\xi}_{i} \big] = \mathbf{0}_{i}$ and covariance matrix $\mathbf{E}\big[ \boldsymbol{\xi}_{i}\boldsymbol{\xi}_{i}^{\intercal} \big] =\Xi_{i}$ with
\begin{align}
\Xi_{i} = 
\begin{bmatrix}
\sigma_{w}^{2} & a\sigma_{w}^{2} & \hdots & a^{i}\sigma_{w}^{2} \\
\star & \sum_{j=0}^{1}a^{2j}\sigma_{w}^{2} & \hdots & \sum_{j=0}^{1}a^{2j+i-1}\sigma_{w}^{2} \\
\vdots & \vdots & \ddots & \vdots \\
\star & \star & \hdots & \sum_{j=0}^{i}a^{2j}\sigma_{w}^{2}
\end{bmatrix} \;. \label{eq:CovMatScalar}
\end{align}

For any $i\in\{1,\cdots,\mathrm{T}\}$, we define the events
\begin{equation}
	\mathcal{F}_{i} = \bigcap_{j=0}^{i-1}\big\{ \vert~\xi_{j}~\vert\leq\epsilon \big\} \label{eqn:Sigma_Set_Scalar}
\end{equation}
with the convention that $\mathcal{F}_{0}$ is a sure event. Thus, we have 
\begin{equation}
	\mathbf{P}(\mathcal{F}_{i}) = F(\epsilon\mathbf{1}_{i}; \mathbf{0}_{i}, \Xi_{i}), \label{eq:ComputeProbability}
\end{equation}
with $\mathbf{P}(\mathcal{F}_{0})=1$.
The transition probabilities can now be computed using the following lemma:
\begin{lemma} \label{lem:Transition_Probabilities}
The transition probabilities $p_{ij}$ in the Markov chain $\{r_{k}\}$, defined in Fig.~\ref{fig:Markov_Chain_wo_Packet_Loss}, are 
\begin{equation*}
p_{ij} = 
\begin{cases}
1-\frac{F(\epsilon\mathbf{1}_{i+1};\mathbf{0}_{i+1},\Xi_{i+1})}{F(\epsilon\mathbf{1}_{i};\mathbf{0}_{i},\Xi_{i})} & \text{if}~i\in\{0,\cdots,\mathrm{T}-1\}, j=0 \\
1 & \text{if}~i=\mathrm{T}, j=0 \\
1-p_{i0} & \text{if}~i\in\{0,\cdots,\mathrm{T}-1\}, j=i+1 \\
0 & \text{otherwise} \;.
\end{cases}
\end{equation*}
\end{lemma}

%
%


A visit of the Markov chain $\{r_{k}\}$ to state $0$ corresponds to a transmission of the control signal from the controller to the actuator, and therefore, in the view of the ergodic theorem,
\begin{align}
	\pi_{0}=\lim\limits_{n\rightarrow\infty}\frac{1}{n}\sum_{k=0}^{n}\mathbbm{1}_{\{r_{k}=0\}}
\end{align}
is the emprical frequency of transmissions.
With the transition probabilities of the Markov chain at hand, we can give an explicit characterization of  the expected communication rate of the event triggered control system:

\begin{theorem}[Communication rate]\label{thm:CommunicationRate_Scalar_wo_Losses}
The expected communication rate between the controller and the actuator for the event-triggered algorithm is given by 
\begin{equation}
	\pi_{0} = \frac{1}{1 + \sum_{n=1}^{\mathrm{T}}\prod_{m=0}^{n-1}(1-p_{m0})} \;. \label{eqn:stationary_distribution_wo_packet_losses}
\end{equation}
\end{theorem}


We are also able to give an explicit characterization of the expected linear-quadratic loss:

\begin{theorem}[Control performance] \label{thm:ControlPerformanceEval_Scalar_wo_Losses}
Consider the problem formulation in Section~\ref{sec:ProblemFormulation} with the event-triggering algorithm and the standard dead-beat controller (i.e., $K=-a/b$). For a given event-threshold $\epsilon> 0$, the empirical average of the control loss function can be computed as
\begin{align}
	J_{\infty} = \Bigg[ \widetilde{Q}\Sigma_{0}^{c} + \sum_{i=1}^{\mathrm{T}} \Big( Q_{x}\Sigma_{i-1} + \widetilde{Q}\Sigma_{i}^{c}\Big)\prod_{j=0}^{i-1}(1-p_{j0})\Bigg]\pi_{0} \;,
	\label{eqn:ControlPerformanceEval_wo_loss}
\end{align}
where $\widetilde{Q}=Q_{x}+\rho K^{\intercal}Q_{u}K$, and 
\begin{align*}
	\begin{array}{lcll}
		\Sigma_{i}        & = & \mathbf{Var}\big[ \xi_{i} \mid \mathcal{F}_{i+1} \big]\;, & i\in [0,\mathrm{T}-1]\;, \\
		\Sigma_{i}^{c}  & = & \mathbf{Var}\big[ \xi_{i} \mid \vert\xi_{i}\vert > \epsilon, \mathcal{F}_{i} \big]\;, & i\in [0,\mathrm{T}-1]\;, \\
		\Sigma_{\mathrm{T}}^{c}  & = & a^{2}\mathbf{Var}\big[ \xi_{\mathrm{T}-1} \mid \mathcal{F}_{\mathrm{T}} \big] + \sigma_{w}^{2}\;. &  \\
	\end{array}
\end{align*}
and $\mathcal{F}_{i}$ defined as in~\eqref{eqn:Sigma_Set_Scalar}.
\end{theorem}

\begin{remark}
	The analytical calculation of the control performance of an event-triggered control system with an arbitrary controller (except dead-beat controller) seems interactable since it is not possible to provide the closed-form of the probability density function of the state $x_{k}$ at any time step $k\in\mathbb{N}_{0}$.
\end{remark}

\begin{remark}
	It is worth noting that the communication rate and control performance are defined by analytical expressions of the mean and variance of truncated random variables. These means and variances do not have explicit closed-form expressions, but can be computed efficiently. 
\end{remark}

Together, these results provide analytic expressions for the communication rate and the emprical average of control loss, for any given threshold $\epsilon$. Next, we investigate how these expressions change when the probability of loss between controller and actuator is non-zero.

\subsection{Control over lossy channel} \label{sec:ETC_FirstOrderSys_w_Loss}
With the basic intuition gained from analyzing the Markov chain that models the loss-free scenario (see Fig.~\ref{fig:Markov_Chain_wo_Packet_Loss}), we now consider the case when the channel between the controller and the actuator exhibits packet loss. 
Recall that if a transmission fails, then the controller will compute a new control command at the next time instant and attempt to transmit this to the actuator. 

%
%
\begin{figure}\centering
	\includegraphics[scale=0.28]{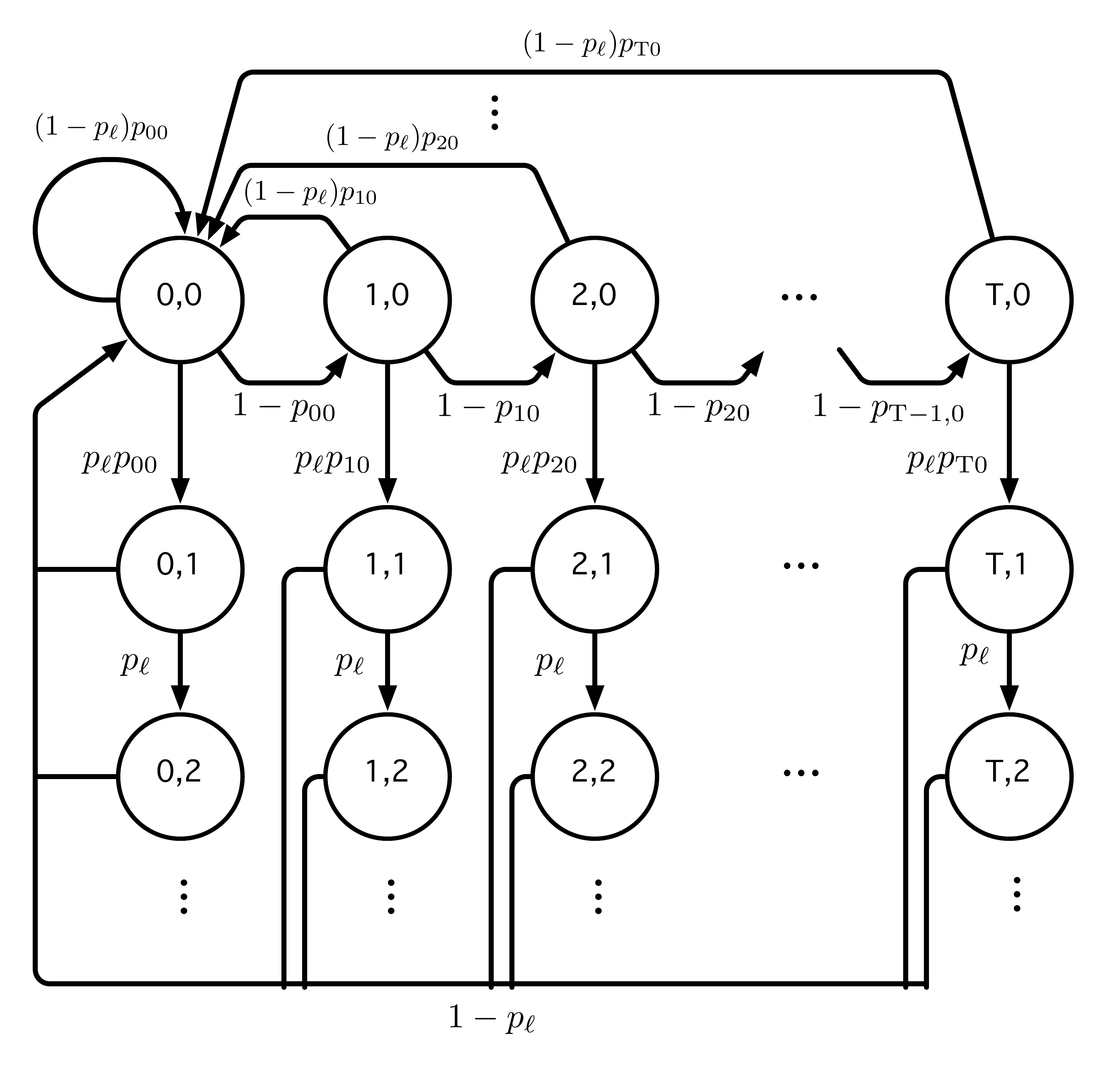}
	\caption{A bidimensional Markov chain $(r_{k},\gamma_{k})$ which illustrates the packet loss in the network.} 
	\label{fig:MarkovChain_with_Packet_Loss}
\end{figure}

Consider the following discrete-time scalar system:
\begin{align}
	x_{k+1} = \phi_{k}(r_{k},\gamma_{k})x_{k} + w_{k}
\end{align}
with initial conditions $x_{0}$, \big(correspondingly $r_{0}\in\{0,1\}$\big) and $\gamma_{0}$, and
\begin{align}
	\phi_{k}(r_{k},\gamma_{k}) = 
	\begin{cases}
		0 & \text{if}~\big\{r_{k}=0~\wedge~\gamma_{k}=0\big\} \\
		a & \text{otherwise}
	\end{cases}
\end{align}
where $\gamma_{k}$ denotes the number of consecutive transmission failures that occurred immediately before time step $k\in\mathbb{N}_{0}$. The process $(r_{k},\gamma_{k})$ evolves according to
\begin{align}
\begin{bmatrix} r_{k+1} \\ \gamma_{k+1} \end{bmatrix} =
\left\lbrace
\begin{array}{ll}
	\hspace*{-3mm}
	\begin{array}{ll}
		\bigl[ \begin{smallmatrix} 0 \\ 0 \end{smallmatrix} \bigr]      & \quad\text{w.p.}~1-p_{\ell} \\
		\bigl[ \begin{smallmatrix} r_{k} \\ 1 \end{smallmatrix} \bigr] & \quad\text{w.p.}~p_{\ell}
	\end{array}
	& \text{if}~\big\{\vert x_{k+1}\vert >\epsilon\wedge\gamma_{k}=0\big\},
    \vspace*{2mm}
	\\
	\vspace*{2mm}
	\hspace*{-3mm}
	\begin{array}{ll}
		\bigl[ \begin{smallmatrix} 0 \\ 0 \end{smallmatrix} \bigr]      & \quad\text{w.p.}~1-p_{\ell} \\
		\bigl[ \begin{smallmatrix} r_{k} \\ 1 \end{smallmatrix} \bigr] & \quad\text{w.p.}~p_{\ell}
	\end{array}
	& \text{if}~\big\{r_{k} >\mathrm{T}\wedge\gamma_{k}=0\big\},
	\\
	\vspace*{2mm}
	\hspace*{-3mm}
	\begin{array}{ll}
		\bigl[ \begin{smallmatrix} 0 \\ 0 \end{smallmatrix} \bigr]      						   & \text{w.p.}~1-p_{\ell} \\
		\bigl[ \begin{smallmatrix} r_{k} \\ \gamma_{k}^{}+1 \end{smallmatrix} \bigr] & \text{w.p.}~p_{\ell}
	\end{array}
	& \text{if}~\gamma_{k}\neq 0, \\
	\hspace*{-1.4mm}
	\bigl[ \begin{smallmatrix} r_{k}^{}+1 \\ 0 \end{smallmatrix} \bigr] & \text{otherwise} \;.
\end{array}
\right. \label{eq:SystemWithPacketLoss_IncStep1}
\end{align}

\begin{lemma}\label{lem:MarkovChain2}
The process $(r_{k}, \gamma_{k}),~k\in\mathbb{N}_{0}$ is an ergodic, time-homogeneous Markov chain in a countably infinite state space $\hat{\mathcal{B}}=\mathcal{B}\times\mathbb{N}_{0}$ with a unique invariant distribution $\boldsymbol{\pi}$.
\end{lemma}

The behavior of the event-triggered communication with packet losses is described by the bidimensional Markov chain $(r_{k},\gamma_{k})$, shown in Fig.~\ref{fig:MarkovChain_with_Packet_Loss}, with state-space $\hat{\mathcal{B}}$. Note that this Markov chain has a countably infinite state-space. The transition probabilities for the Markov chain depend on the packet loss probability $p_{\ell}$ and the transition probabilities already characterized in Lemma~\ref{lem:Transition_Probabilities} for the loss-free case. Thus, we now have all information necessary to derive the expected rate of attempted and successful transmissions from the Markov chain.


\begin{theorem}[Communication rate] \label{thm:CommunicationRate_Scalar_with_Losses}
	Under the event triggering mechanism (defined in~\S\ref{sec:ControllerDesign}), the expected rate of the successful reception of control packets at the actuator is calculated as
	\begin{align}
		\pi_{00} = \frac{1}{\Theta_{0} + \sum_{n=1}^{\mathrm{T}}\Theta_{n}\prod_{m=0}^{n-1}(1-p_{m0}) },	\label{eqn:stationary_distribution_w_packet_losses}
	\end{align}
	where $\Theta_{i}=\frac{1-p_{\ell}(1-p_{i0})}{1-p_{\ell}},~\forall i\in\{0,\cdots,\mathrm{T}\}$, and the expected rate of attempted transmissions between the controller and actuator is obtained as 
	\begin{align}
		\widetilde{\pi}_{00} = \frac{1}{1 + \sum_{n=1}^{\mathrm{T}}\prod_{m=0}^{n-1}(1-p_{m0})} \;. \label{eqn:stationary_distribution_attempted}
	\end{align}
\end{theorem}


\begin{remark}
Note that we consider a discrete-time stochastic system whose underlying Markov chain has a countably infinite state-space. Typically, the solution of optimal control problems for such systems rely on the study of a countably infinite set of coupled Riccati equations (e.g., ~\cite{FrB:02, FrB:06}) and can only offer numerical approximations of the long-run average cost. In contrast, our framework with the dead-beat control assumption yields an anlytical characterization of the control performance that allows for efficient numerical computations.
\end{remark}

\begin{theorem}[Control performance] \label{thm:ControlPerformanceEval_Scalar_with_Losses}
Consider the problem formulation in Section~\ref{sec:ProblemFormulation} with the event-triggering algorithm and the dead-beat controller described in Section~\ref{sec:ControllerDesign}. 
Suppose that
%
\begin{align*}
	p_{\ell}a^{2} < 1 \;,
\end{align*}
Then, for a given event-threshold $\epsilon>0$, the empirical average of the control loss function is 
%
\begin{align}
	J_{\infty} =\; \Bigg[ p_{00}\widehat{Q}\bigg( \Sigma_{0}^{c} + \frac{p_{\ell}\sigma_{w}^{2}}{1-p_{\ell}} \bigg) + \sum_{i=1}^{\mathrm{T}}\bigg[Q_{x}\Sigma_{i-1} + p_{i0}\widehat{Q}\bigg( \Sigma_{i}^{c}+\frac{p_{\ell}\sigma_{w}^{2}}{1-p_{\ell}}\bigg) \bigg]\prod_{j=0}^{i-1}(1-p_{j0}) \Bigg]\pi_{00},
	\label{eqn:ControlLoss_LossyNetworks}
\end{align}
%
where $\widehat{Q}=\frac{(1-p_{\ell})( Q_{x}+\rho K^{\intercal}Q_{u}K )+p_{\ell}Q_{x}}{1-p_{\ell}a^{2}}$ and the truncated variances:
\begin{align*}
	\begin{array}{lcll}
		\Sigma_{i}        & = & \mbox{Var}\big[ \xi_{i} \mid \mathcal{F}_{i+1} \big]\;, & i\in[0, \mathrm{T}-1]\;, \\
		\Sigma_{i}^{c}  & = & \mbox{Var}\big[ \xi_{i} \mid \vert\xi_{i}\vert > \epsilon, \mathcal{F}_{i} \big]\;, & i\in[0, \mathrm{T}-1 ]\;, \\
		\Sigma_{\mathrm{T}}^{c}  & = & a^{2}\mbox{Var}\big[ \xi_{\mathrm{T}-1} \mid \mathcal{F}_{\mathrm{T}} \big] + \sigma_{w}^{2}\;, &  \\
	\end{array}
\end{align*}
and $\mathcal{F}_{i}$ defined as in~\eqref{eqn:Sigma_Set_Scalar}.
\end{theorem}

\begin{remark}
Notice that the condition $p_{\ell}a^{2}<1$ is identical to the necessary and sufficient condition for stability of scalar systems under periodic control over packet drop channels; see~\emph{e.g.},~\cite{ZBP:01}. Accordingly, this condition must be satisfied to compute a finite control loss for our set-up.
\end{remark}

\section{Event-triggered control of higher-order systems} \label{sec:MainResultsHigherOrder}

We present our result in two steps. Section~\ref{sec:ETC_HighOrderSys} outlines the results for the vector case with arbitrary $n$ when packet loss $p_{\ell}=0$. We extend our analysis to the case with packet dropout in Section~\ref{sec:ETC_HighOrderSys_w_Loss}. 

\subsection{Control over perfect channel}\label{sec:ETC_HighOrderSys}

In this subsection, we extend our results to the case when the process state $\x_{k}$ is a vector. This is not a trivial extension. Unlike the scalar case, the control packets now need to be stored in a buffer at the actuator. Even when there are no packet losses, the one dimensional process $\{r_{k}\}$ becomes a bidimensional process $(r_{k},\eta_{k})$. At each time instant $k\in\mathbb{N}_{0}$, the controller $\mathcal{C}$ uses the state $\x_{k}$ of the system~\eqref{eqn:SystemModel_VectorCase} to calculate and send a packet with a sequence of control commands as described in~\eqref{eqn:ControlSequence}. We can rewrite the evaluation of the process $\{\x_{k}\}_{k\in\mathbb{N}_{0}}$ of~\eqref{eqn:SystemModel_VectorCase} as follows:
\begin{align}
	\begin{bmatrix} \x_{k+1} \\ \eta_{k+1} \end{bmatrix} = 
	\begin{cases}
		A\x_{k} + \w_{k} & \text{if}~\big\{\parallel\x_{k}\parallel_{\infty}\leq\epsilon\wedge\eta_{k}=0\big\}\;, \\
		\bphi_{k}\x_{k-\eta_{k}^{}+1}+\sum\limits_{l=0}^{\eta_{k}^{}-1}A^{l}\w_{k-l} & \text{otherwise}\;,
	\end{cases}
\end{align}
where
\begin{align*}
	\bphi_{k} = 
	\begin{cases}
		(A+BK)^{\nu-\eta_{k}^{}}, & \text{if}~\eta_{k}\in\{ 1, \cdots, \nu-1\}\;, \\
		\mathbf{0}_{n}, & \text{if}~\eta_{k} = 0\;,
	\end{cases}
\end{align*}
and $\eta_{k}$ denotes the number of control packets in the buffer.

The evolution of the process pair $(r_{k},\eta_{k})$ is defined by
\begin{align}
\begin{bmatrix} r_{k+1} \\ \eta_{k+1} \end{bmatrix} =
\left\lbrace
\begin{array}{ll}
	\bigl[ \begin{smallmatrix} 0 \\ \nu-1 \end{smallmatrix} \bigr]
	& 
	\text{if}~\big\{\parallel x_{k+1}\parallel_{\infty} >\epsilon~\wedge~\eta_{k}=0\big\},
	\\
	\bigl[ \begin{smallmatrix} 0 \\ \nu-1 \end{smallmatrix} \bigr]
	&
	\text{if}~\big\{r_{k} >\mathrm{T}~\wedge~\eta_{k}=0\big\},
	\\
	\bigl[ \begin{smallmatrix} 0 \\ 0 \end{smallmatrix} \bigr]
	& \text{if}~\eta_{k} = 1,
	\\
	\bigl[ \begin{smallmatrix} r_{k} \\ \eta_{k}^{}-1 \end{smallmatrix} \bigr]
	& \text{if}~\eta_{k}\in\{2,\cdots,\nu-1\}
	\\
	\bigl[ \begin{smallmatrix} r_{k}^{}+1 \\ 0 \end{smallmatrix} \bigr] & ~\text{otherwise} \;.
\end{array}
\right.
\end{align}

\begin{lemma}\label{lem:MarkovChain3}
The process $(r_{k}, \eta_{k}),~k\in\mathbb{N}_{0}$ is an ergodic, time-homogeneous Markov chain with a state space $\tilde{\mathcal{B}}=\mathcal{B}\times\{0,\cdots,\nu-1\}$ and it has a unique invariant distribution~$\boldsymbol{\pi}$. 
\end{lemma}

\begin{figure}\centering
	\includegraphics[scale=0.28]{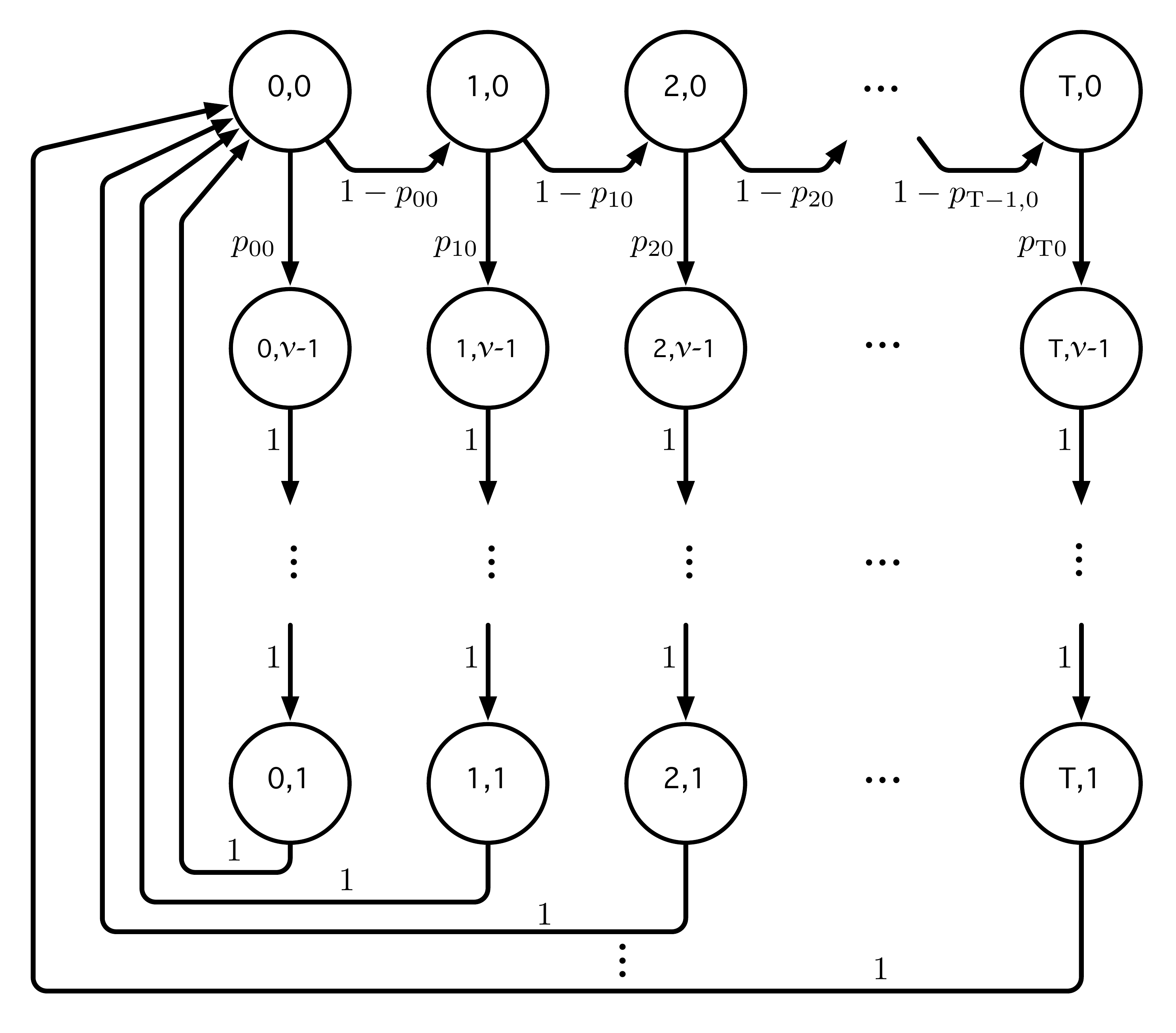}
	\caption{A bidimensional Markov chain $(r_{k},\eta_{k})$ that illustrates the event-triggered transmission for high-order systems.} 
	\label{fig:MarkovChain_VectorCase}
\end{figure}

The behavior of the event-triggered communication of the system~\eqref{eqn:SystemModel_VectorCase} with the packetized controller~\eqref{eqn:ControlSequence} can be described by Markov chain $(r_{k},\eta_{k})$ (see Fig.~\ref{fig:MarkovChain_VectorCase}). This discrete-state discrete-time Markov chain has $(\mathrm{T}+1)\times\nu$ states\footnote{Recall that $\nu$ represents the controllability index of the system~\eqref{eqn:SystemModel_VectorCase}.}. Each mode is represented by two digits: the first digit denotes when the last transmission took place, and the second one denotes how many control packets are used from the buffer.

%
%

To describe how the transition probabilities depend on the system parameters and the event threshold, we define the random variables, i.e.,
\begin{align*}
	\bdel_{i,k} = 
	\begin{cases}
		\sum_{l=0}^{\nu-1}A^{l}w_{k-l} & \text{if}~i=0\;, \\
		\sum_{l=0}^{\nu-1}A^{l+i}w_{k-l} + \sum_{j=0}^{i-1}A^{j}w_{k+i-j} & \text{if}~i\neq 0\;.
	\end{cases}
\end{align*}
for all $i\in\{0,1,\cdots, \mathrm{T}-1\}$. Note that we only consider the random variables that describe the evolution of states when there is no transmission, and when we reset the mean of the state because the rest of the events are sure events -- the transition probabilities are equal to unity.

Unlike the random variable $\xi_{i,k}$ (defined in~\S~\ref{sec:ETC_FirstOrderSys}), $\bdel_{i,k}$ is vector-valued. However, the probability density function of $\bdel_{i,k}$ is also time-invariant since the noise process $\{\w_{k}\}_{k\in\mathbb{N}_{0}}$ is white and stationary. We thus neglect the time index to simplify notation and denote $\bdel_{i,k}$ by $\bdel_{i}$. Similar to the scalar case, the augmented random variable $\Delta_{in}\triangleq \big[ \bdel_{0}~\bdel_{1}~\cdots~\bdel_{i} \big]^{\intercal}$ 
a multi-variate normal distribution with mean $\mathbf{E}\big[ \Delta_{in} \big] = \mathbf{0}_{in}$ and covariance matrix $\mathbf{E}\big[ \Delta_{in}\Delta_{in}^{\intercal} \big] =\Xi_{in}$ given by
\begin{align*}
	\begin{small}
	\begin{bmatrix}
		\Sigma^{*} & \Sigma^{*}A^{\intercal} & \hdots & \Sigma^{*}(A^{\intercal})^{i} \\
		\star & A\Sigma^{*}A^{\intercal} + \Sigma_{w} & \hdots & A\Sigma^{*}(A^{\intercal})^{i}  + \Sigma_{w}(A^{\intercal})^{i-1}  \\
		\vdots & \vdots & \ddots & \vdots \\
		\star & \star & \hdots & A^{i}\Sigma^{*}(A^{\intercal})^{i} + \sum\limits_{j=0}^{i-1}A^{j}\Sigma_{w}(A^{\intercal})^{j} 
	\end{bmatrix} 
	\end{small}
\end{align*}
where $\Sigma^{*}\triangleq\sum_{l=0}^{\nu-1}A^{l}\Sigma_{w}A^{\intercal l}$. As defined in~\eqref{eqn:Sigma_Set_Scalar}, we establish the events, for all $i\in\{1,\cdots,\mathrm{T}\} $,
\begin{equation}
	\mathcal{F}_{i} = \bigcap_{j=0}^{i-1}\big\{ \parallel\bdel_{j}\parallel_{\infty}\leq\epsilon \big\}\;, \label{eqn:Sigma_Set_Vector}
\end{equation}
where $\mathcal{F}_{0}$ is a sure event.
For the transition probabilities, we use the shorthand notation:
%
\begin{equation*}
	\mathbf{P}\big( (i_{1},j_{1}) \mid (i_{0},j_{0}) \big)\triangleq \mathbf{P}\big( (r_{k+1},\eta_{k+1})=(i_{1},j_{1}) \mid (r_{k},\eta_{k})=(i_{0},j_{0}) \big).
\end{equation*}
%

The transition probabilities can now be computed using the following lemma.

\begin{lemma} \label{lem:Transition_Probabilities_Vector}
The non-null transition probabilities of the Markov chain $(r_{k},\eta_{k}),~k\in\mathbb{N}_{0}$, shown in Fig.~\ref{fig:MarkovChain_VectorCase}, are
\begin{align*}
\begin{array}{lcll}
	\mathbf{P}\big( (0, \nu-1) \mid (0, 0) \big) &=& p_{0}\;, & \\
	\mathbf{P}\big( (1, 0) \mid (0, 0) \big) &=& 1-p_{0}\;, & \\
	\mathbf{P}\big( (i, \nu-1) \mid (i, 0) \big) &=& p_{i}\;, & i\in[1,\mathrm{T}-1] \;, \\
	\mathbf{P}\big( (i+1, 0) \mid (i, 0) \big) &=& 1 - p_{i}\;, & i\in[1,\mathrm{T}-1] \;, \\
	\mathbf{P}\big( (\mathrm{T},\nu-1) \;\vert\; (\mathrm{T},0) \big) &=& 1\;, &  \\
	\mathbf{P}\big( (i,j-1) \;\vert\; (i,j) \big) &=& 1\;, & i\in[0,\mathrm{T}],~~j\in[1,\nu-1] \;, 
\end{array}
\end{align*}
where
\begin{align*}
	p_{i} = 1-\frac{F(\epsilon\mathbf{1}_{(i+1)n};\mathbf{0}_{(i+1)n},\Xi_{(i+1)n})}{F(\epsilon\mathbf{1}_{in};\mathbf{0}_{in},\Xi_{in})}, \quad \forall i\in[0,\mathrm{T}-1] \;.
\end{align*}
\end{lemma}

We now proceed to characterize the analytical model of the communication frequency in case the communication between the controller and the actuator.


\begin{theorem}[Communication rate]\label{thm:CommunicationRate_Vector}
	The expected communication rate between the controller and the actuator for the event-triggered algorithm is obtained as
	\begin{align}
		\pi_{00} = \frac{1}{\Theta_{0}+\sum_{i=1}^{\mathrm{T}}\Theta_{i}\prod_{j=0}^{i-1}(1-p_{j})},
		\label{eqn:CommunicationRate_HighOrderSys}
	\end{align}
	with $\Theta_{i} = 1+(\nu-1)p_{i},~\forall i\in\{0,\cdots.\mathrm{T}\}$.
\end{theorem}

The next theorem characterizes the expected linear-quadratic loss:
\begin{theorem}[Control performance] \label{thm:ControlPerformanceEval}
Consider the problem formulation in Section~\ref{sec:ProcessModel_HOS} with the event-triggering algorithm and the control strategy described in Section~\ref{sec:ControllerDesign}. For a given event-threshold $\epsilon> 0$, the empirical average of the linear-quadratic control loss is computed as
%
%
\begin{align}
	J_{\infty} =\; \Bigg[p_{0}\big(  X_{0} + Z  \big) + \sum_{i=1}^{\mathrm{T}}\bigg[ \mbox{Tr}\big( Q_{x}\Sigma_{i-1} \big)
	+ p_{i}\big( X_{i} + Z \big) \bigg]\prod_{l=0}^{i-1}(1-p_{l})\Bigg]\pi_{00} \;,
	 \label{eqn:ControlLoss_VectorCase}
\end{align}
where, for all $i\in \{0,\cdots, \mathrm{T}\}$,
\begin{align*}
	X_{i}=&\; \sum_{j=0}^{\nu-1}\mbox{Tr}\Big( (A_{c}^{\intercal})^{j}\widetilde{Q}A_{c}^{j}\Sigma_{0}^{c} \Big)\;, \\
	Z =&\; \sum_{j=0}^{\nu-2}\sum_{l=0}^{j}\mbox{Tr}\Big( (A^{\intercal})^{l}Q_{x}A^{l}\Sigma_{w} \Big)\;,
\end{align*}
%
with $\widetilde{Q}=Q_{x}+\rho K^{\intercal}Q_{u}K$, $A_{c}=A+BK$, and the truncated covariances:
\begin{align*}
	\begin{array}{lcll}
		\Sigma_{i}        & = & \mathbf{Cov}\big[ \boldsymbol{\delta}_{i} \mid \mathcal{F}_{i+1} \big]\;, & i\in [0,\mathrm{T}-1]\;, \\
		\Sigma_{i}^{c}  & = & \mathbf{Cov}\big[ \boldsymbol{\delta}_{i} \mid \parallel\boldsymbol{\delta}_{i}\parallel_{\infty} > \epsilon, \mathcal{F}_{i} \big]\;, & i\in [0,\mathrm{T}-1]\;, \\
		\Sigma_{\mathrm{T}}^{c}  & = & A\mathbf{Cov}\big[ \boldsymbol{\delta}_{\mathrm{T}-1} \mid \mathcal{F}_{\mathrm{T}} \big]A^{\intercal} + \Sigma_{w}\;. &  \\
	\end{array}
\end{align*}
and $\mathcal{F}_{i}$ defined as in~\eqref{eqn:Sigma_Set_Vector}.
\end{theorem}

The aforementioned results allow one to obtain analytical expressions for the communication rate and the empirical average of the control loss for any given event-threshold $\epsilon> 0$.

\subsection{Control over lossy channel} \label{sec:ETC_HighOrderSys_w_Loss}
If the state vector $\x_{k}$ is transmitted over the lossy channel between the controller and the actuator, the behavior of the event-triggered communication with packet losses is described by the three-dimensional Markov chain $(r_{k},\gamma_{k},\eta_{k})$, $k\in\mathbb{N}_{0}$ with countably infinite state space. The Markov chain can be obtained from the one depicted in Fig.~\ref{fig:MarkovChain_VectorCase} using the same technique that was used to generalize the Markov chain of \S~IV-A to that of \S~IV-B. To compute the transition probabilities for this Markov chain, it is convenient to use Lemma~\ref{lem:Transition_Probabilities_Vector}, derived in the previous subsection. In fact, we have all necessary information to derive the expected rate of successful transmissions because the packet-loss probability $p_{\ell}$ is the only additonal parameter.

\begin{theorem}[Communication rate] \label{thm:CommRate_Vector_Loss}
Under the event triggered mechanism (defined in~\S\ref{sec:ControllerDesign}), the expected rate of successful reception of control packets at the actuator node is obtained as
	\begin{align}
		\pi_{000} = \frac{1}{\Theta_{0}+\sum_{i=1}^{\mathrm{T}}\Theta_{i}\prod_{j=0}^{i-1}(1-p_{j})}, \label{eqn:CommRate_Vector_Loss}
	\end{align}
	with $\Theta_{i} = 1+(\nu-1)p_{i}+\frac{p_{\ell}p_{i}}{1-p_{\ell}},~\forall i\in\{0,\cdots.\mathrm{T}\}$, and the communication rate between the controller and the actuator is computed as
	\begin{align}
		\widetilde{\pi}_{000} = \frac{1}{\Theta_{0}+\sum_{i=1}^{\mathrm{T}}\Theta_{i}\prod_{j=0}^{i-1}(1-p_{j})}, 
	\end{align}
	with $\Theta_{i} = 1+(\nu-1)p_{i},~\forall i\in\{0,\cdots.\mathrm{T}\}$.
\end{theorem}

The next theorem characterizes the expected linear-quadratic cost.

\begin{theorem}[Control performance] \label{thm:CntrlPerf_Vector_Loss}
Consider the problem formulation in Section~\ref{sec:ProblemFormulation} with the event-triggering algorithm and the dead-beat controller described in Section~\ref{sec:ControllerDesign}. 
Suppose that 
\begin{align*}
	p_{\ell}\lambda_{\max}(A)^{2} < 1\;.
\end{align*}
Then, for a given event-threshold $\epsilon> 0$, the empirical average of the control loss function is 
%
%
\begin{multline}
	J_{\infty} = \Bigg[p_{0}\bigg[ \sum_{j=0}^{\nu-1}X_{0j} + \frac{p_{\ell}Y_{0}}{1-p_{\ell}} + Z \bigg] + \sum_{i=1}^{\mathrm{T}}\bigg[ \mbox{Tr}\big( Q_{x}\Sigma_{i-1} \big) \\
	 + p_{i}\Big( \sum_{j=0}^{\nu-1}X_{ij} + \frac{p_{\ell}Y_{i}}{1-p_{\ell}} + Z\Big)\bigg]\prod_{l=0}^{i-1}(1-p_{l})\Bigg]\pi_{000} \;, \label{eqn:ControlLoss_VectorCase_Loss}
\end{multline}
where
\begin{align*}
	X_{ij} =&\; \mbox{Tr}\Big( \Theta_{j}\big( (1-p_{\ell})\Sigma_{i}^{c} + p_{\ell}\Sigma_{w} \big) \Big)\;, \\
	Y_{i} =&\; \mbox{Tr}\Big( \Upsilon\big( (1-p_{\ell})\Sigma_{i}^{c} + p_{\ell}\Sigma_{w} \big) \Big)\;, \\
	Z =&\; \sum_{j=0}^{\nu-2}\sum_{l=0}^{j}\mbox{Tr}\Big( (A^{\intercal})^{l}Q_{x}A^{l}\Sigma_{w} \Big)\;, 
\end{align*}
with, for all $j\in \{0,\cdots, \nu-1\}$,
\begin{align*}
	\Upsilon =&\; \mathtt{lyap}\big(\sqrt{p_{\ell}}A^{\intercal},Q_{x}\big)\;, \\
	\Theta_{j} =&\; \mathtt{lyap}\big(\sqrt{p_{\ell}}A^{\intercal},A_{c}^{\intercal j}\widetilde{Q}A_{c}^{j}\big)\;,
\end{align*}
where $\widetilde{Q}=Q_{x}+\rho K^{\intercal}Q_{u}K$, $A_{c}=A+BK$ and the truncated variances:
\begin{align*}
	\begin{array}{lcll}
		\Sigma_{i}        & = & \mathbf{Cov}\big[ \boldsymbol{\delta}_{i} \mid \mathcal{F}_{i+1} \big]\;, & i\in [0,\mathrm{T}-1]\;, \\
		\Sigma_{i}^{c}  & = & \mathbf{Cov}\big[ \boldsymbol{\delta}_{i} \mid \parallel\boldsymbol{\delta}_{i}\parallel_{\infty} > \epsilon, \mathcal{F}_{i} \big]\;, & i\in [0,\mathrm{T}-1]\;, \\
		\Sigma_{\mathrm{T}}^{c}  & = & A\mathbf{Cov}\big[ \boldsymbol{\delta}_{\mathrm{T}-1} \mid \mathcal{F}_{\mathrm{T}} \big]A^{\intercal} + \Sigma_{w}\;. &  \\
	\end{array}
\end{align*}
and $\mathcal{F}_{i}$ defined as in~\eqref{eqn:Sigma_Set_Vector}.
\end{theorem}

\begin{remark}
Note once again that the condition $p_{\ell}\lambda_{\max}(A)^{2}<1$ is a natural assumption needed for the stabilization of the system~\eqref{eqn:SystemModel_VectorCase} by any control algorithm when the controller-actuator communication is across an unreliable link with the erasure probability $p_{\ell}\in (0,1)$; see e.g.,~\cite{ZBP:01}.
\end{remark}

\section{Numerical Examples} \label{sec:NumExamples}
We will now illustrate how the analysis techniques developed in Section~\ref{sec:MainResults} allow us to study the trade-off between the communication rate and the closed-loop performance for the event-triggered scheme detailed in Section~\ref{sec:ProblemFormulation}.

\subsection{Event-triggered control for first-order systems} \label{sec:Example_Scalar_no_Loss}
We first consider a scalar linear stochastic system
\begin{align*}
	x_{k+1} = 1.6x_{k} + u_{k} + w_{k} \;,
\end{align*}
where $\{w_{k}\}_{k\in\mathbb{N}_{0}}$ is a zero-mean Gaussian white noise process with variance $\sigma_{w}^2=1.44$. The initial state $x_{0}$ has a normal distribution with zero mean and variance $\sigma_{0}^{2}=1.0$. We consider a cheap control scenario, where $Q_{x}=Q_{u}=1$ and  $\rho\rightarrow 0$, whose optimal control under periodic sampling is the dead-beat control law $u_{k}=-1.6x_{k}$. The number of sample times between two consecutive transmissions is upper-bounded by the time-out interval $\mathrm{T}=5$.
%
%
\begin{figure}\centering
	\includegraphics[scale=1.0]{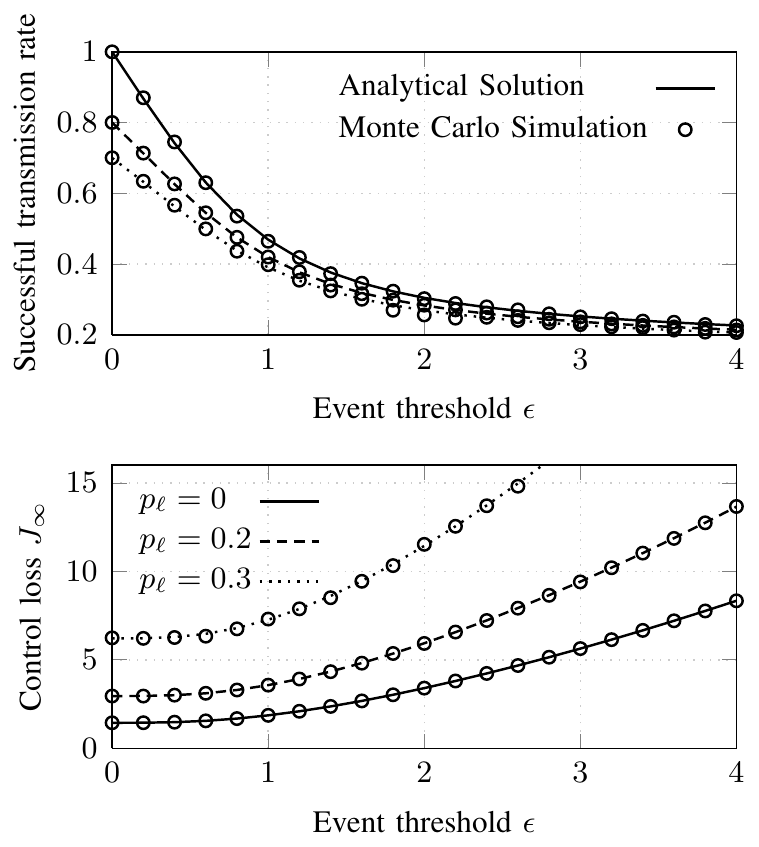}
	\caption{A comparison of the successful reception rate (resp. control performance) obtained from the analytic expressions~\eqref{thm:CommunicationRate_Scalar_wo_Losses} and~\eqref{thm:CommunicationRate_Scalar_with_Losses} (resp.~\eqref{thm:ControlPerformanceEval_Scalar_wo_Losses} and~\eqref{thm:ControlPerformanceEval_Scalar_with_Losses}) and Monte Carlo simulations for several packet loss probabilities; e.g., $p_{\ell}\in\{ 0, 0.2, 0.3 \}$. The marked curve with "$\circ$" is obtained by averaging $10,000$ Monte Carlo simulations for the horizon length $25,001$ samples with the process noise $\{w_{k}\}_{k\in\mathbb{N}_{0}}$ and the initial condition $x_{0}$ generated randomly.} 
  \label{fig:ETC_Performence__LossyNetworks}
\end{figure}

In Fig.~\ref{fig:ETC_Performence__LossyNetworks}, we examine the communication rate and the control loss  predicted by Theorem~\ref{thm:CommunicationRate_Scalar_wo_Losses} and~\ref{thm:ControlPerformanceEval_Scalar_wo_Losses} in the absence of packet loss (i.e., $p_{\ell}=0$) for different thresholds $\epsilon\geq 0$. We first note how the analytical results (solid line) matches the Monte Carlo simulations (circles) almost perfectly.  

Next, we observe that changing the threshold from $0$ (periodic control) to $1$ yields significant reduction in the communication frequency at the expense of a small increase in the control loss. On the other hand, increasing the threshold from $2$ to $4$ only brings little (absolute) reduction in communication, but causes a large degradation in the closed-loop performance.
Such observations may be useful for designing the event-triggering scheme.

Fig.~\ref{fig:ETC_Performence__LossyNetworks} also shows how the successful reception rate and the control loss are impacted by a lossy communication channel (Theorem~\ref{thm:CommunicationRate_Scalar_with_Losses} and~\ref{thm:ControlPerformanceEval_Scalar_with_Losses}). Again, the analytical model has nearly perfect correspondance with Monte Carlo simulations. Moreover, we note that loss rates up until $p_{\ell}=0.2$ have have only a small effect on the communication rate and closed-loop performance, while $p_{\ell}=0.3$ leads to a specified performance loss compared to the case of reliable transmission. Recall that if the packet loss probability is larger than  $p_{\ell}^{\star}=1/1.6^{2}\approx 0.3906$, the event-triggered control system will be unstable.
%
%
\begin{figure}\centering
	\includegraphics[scale=1.0]{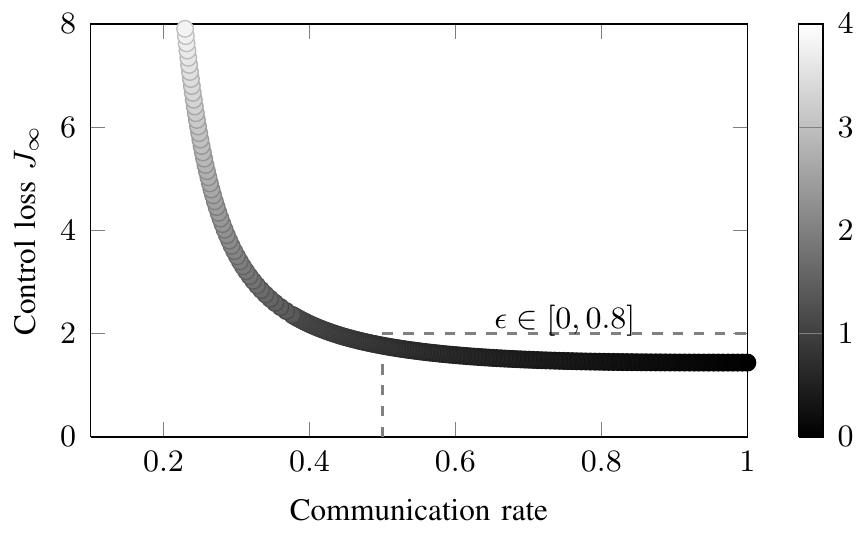}
	\caption{The control performance for different communication frequency and the event thresholds (shown in gray scale).} 
   \label{fig:Tradeoff_Curve}
\end{figure}

An alternative visualization of the trade-off between control loss and communication rate is shown in Figure~\ref{fig:Tradeoff_Curve}. Different points on the curve correspond to different even thresholds ranging from $0$ to $4$ and communication rates and control losses are calculated using Theorem~\ref{thm:CommunicationRate_Scalar_wo_Losses} and Theorem~\ref{thm:ControlPerformanceEval_Scalar_wo_Losses}, respectively. Note that the communication rate decreases dramatically with an increased control loss as the threshold $\epsilon$ varies between $0$ and $2.25$ (dark colors). For $\epsilon>2.25$ (lighter color), both quantities become less sensitive to changes in the threshold value. We can also identify $\epsilon\in [0,0.8]$ as a particularly attractive region where a large decrease in the communication rate can be obtained for a small loss in control performance.

\subsection{Event-triggered control for high-order systems}
Next, we consider a second-order plant with state-space representation
\begin{align*}
	\x_{k+1} = \begin{bmatrix} 2.2 & -1.2 \\ 1 & 0 \end{bmatrix}\x_{k} + \begin{bmatrix} 0.8 \\ 0.4 \end{bmatrix}u_{k} + \w_{k} \;.
\end{align*}
For the periodic cheap control with $Q_x=\bigl[\begin{smallmatrix} 1 & 0 \\ 0 & 1 \end{smallmatrix} \bigr]$, $Q_u=1$ and $\rho\rightarrow 0$, the optimal controller is 
\begin{align*}
	u_{k} = \begin{bmatrix} -\frac{19}{8} & -\frac{3}{4} \end{bmatrix}\x_{k} \;.
\end{align*}
This control is of dead-beat type and drives any system state to the origin in two time steps. The consider the initial condition $\x_{0}$ as a Gaussian random variable with zero mean and covariance $\Sigma_{0} = \bigl[\begin{smallmatrix} 6.224 & 2.16 \\ 2.16 & 2 \end{smallmatrix} \bigr]$, while the process noise $\{\w_{k}\}_{k\in\mathbb{N}_{0}}$ is white, zero mean, Gaussian with covariance $\Sigma_{w} = \bigl[\begin{smallmatrix} 1 & 0.2 \\ 0.2 & 1 \end{smallmatrix} \bigr]$. The process noise and the initial condition are mutually independent. The time-out interval is set to $\mathrm{T}=4$.
%
%
\begin{figure}\centering
	\includegraphics[scale=1.0]{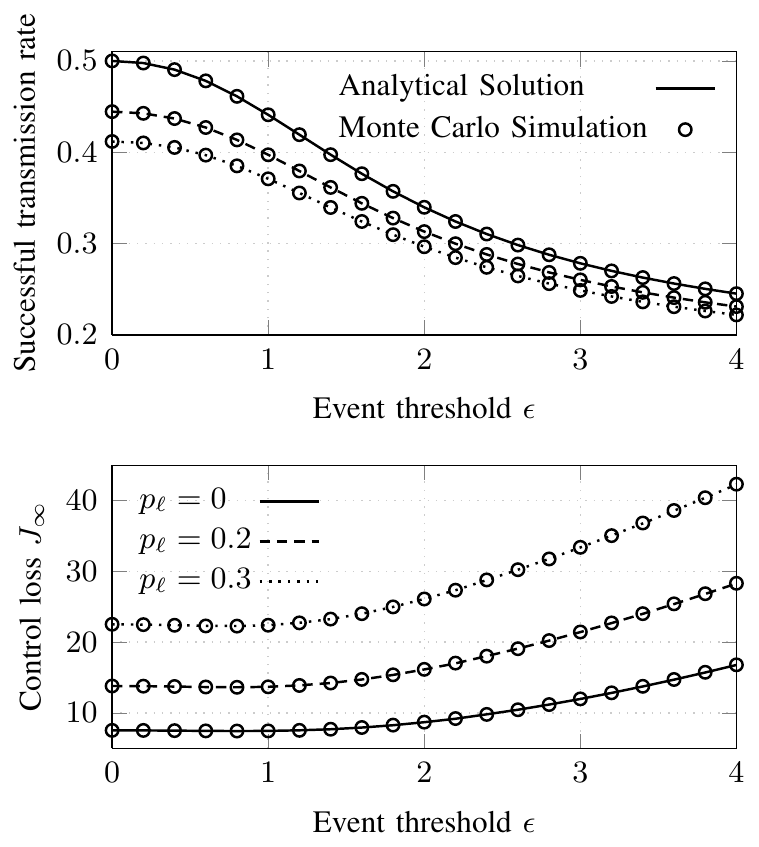}
	\caption{A comparison of the successful reception rate (resp. control performance) obtained from the analytic expressions~\eqref{eqn:CommunicationRate_HighOrderSys} and~\eqref{eqn:CommRate_Vector_Loss} (resp.~\eqref{eqn:ControlLoss_VectorCase} and~\eqref{eqn:ControlLoss_VectorCase_Loss}) and Monte Carlo simulations for several packet loss probabilities; e.g., $p_{\ell}\in\{ 0, 0.2, 0.3 \}$. The marked curve with "$\circ$" is obtained by averaging $10,000$ Monte Carlo simulations for the horizon length $25,001$ samples with the process noise $\{\w_{k}\}_{k\in\mathbb{N}_{0}}$ and the initial condition $\x_{0}$ generated randomly.} 
  	\label{fig:ETC_Vector}
\end{figure}

As shown in Figure~\ref{fig:ETC_Vector}, the analytical results obtained in Theorem~\ref{thm:CommunicationRate_Vector} and~\ref{thm:ControlPerformanceEval} match Monte Carlo simulations perfectly. Note that the successful communication rate is limited to $0.5$ since $\nu=2$. Therefore, the event-triggered control sends at most one packet every two sampling instances. Apart from this, the behaviour is qualitatively similar to the scalar case.

\begin{figure}\centering
	\includegraphics[scale=1.0]{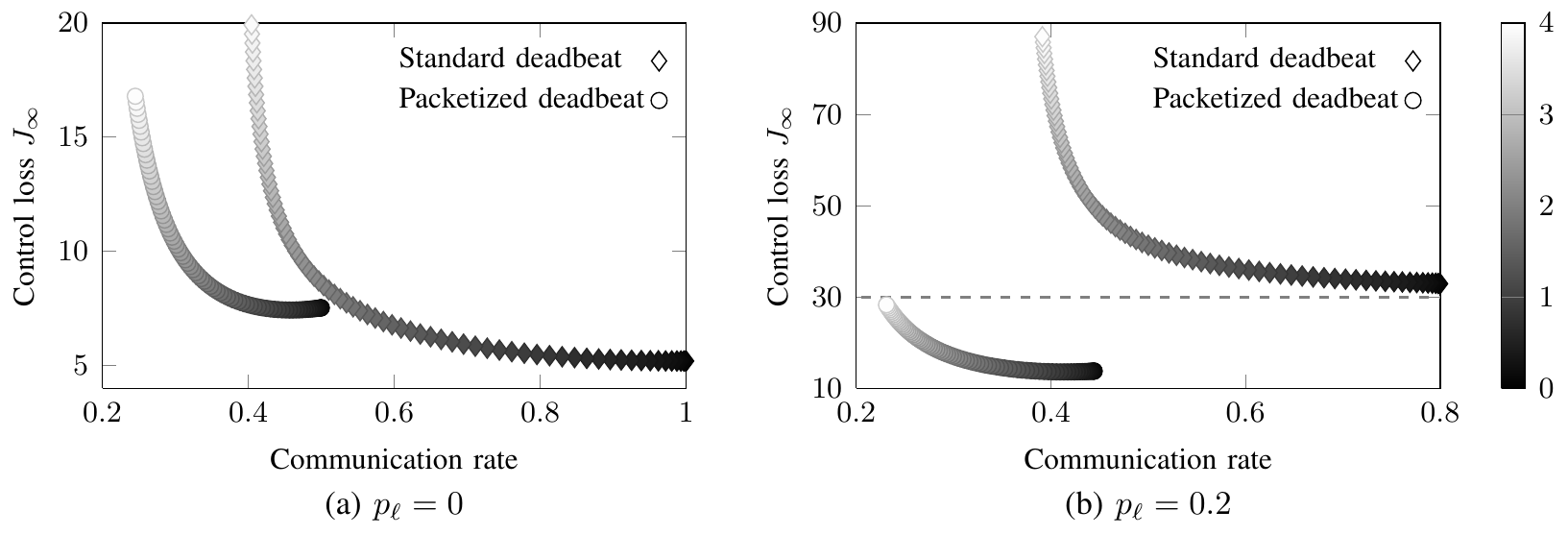}
	\caption{A comparison of the communication rate and the control performance of the event-triggered control system with and without packetized dead-beat controller for two packet loss rates: (a) $p_{\ell}=0$ and (b) $p_{\ell}=0.2$ (event threshold is shown with gray scale). The marked curves with "$\circ$" and "$\diamond$" are obtained by averaging $10,000$ Monte Carlo simulations for the horizon length $25,001$ samples with the process noise $\{\w_{k}\}_{k\in\mathbb{N}_{0}}$ and the initial condition $\x_{0}$ generated randomly.} 
  	\label{fig:ETC_Vector_Comp}
\end{figure}

Figure~\ref{fig:ETC_Vector_Comp}(a) highlights the differences between our packetized event-based control algorithm and \emph{the threshold-based event-triggered control algorithm}, formed as
\begin{align*}
	u_{k} = 
	\begin{cases}
		K\x_{k} & \text{if}~\parallel \x_{k}\parallel_{\infty}>\epsilon \;, \\
		0			& \text{otherwise} \;.
	\end{cases} 
\end{align*} 
For the lossless case ($p_{\ell}=0$), Figure~\ref{fig:ETC_Vector_Comp}(a) shows that the packetized event-triggered control outperforms the standard threshold-based implementation as long as the average successful transmission rate is smaller than $0.5$. If  we allow for higher communication rates, then the standard implementation performs better on average.
%
%

The differences between the two variations are more striking in the case of packet losses. Figure~\ref{fig:ETC_Vector_Comp}(b) compares communication vs control trade-off for the packetized and the standard implementation of the threshold-based event-triggered control when the loss rate is $20\%$. In this case, the packetized implementation strictly dominates the standard implementation, and when a comparable performance is searched for, this can be done at a dramatic decrease in communication cost. This performance improvement can be understood by observing that to reset the state $\x_{k}$ in absence of the process noise, it is neccessary to apply two consecutive control commands computed by the dead-beat controller. Whenever the packetized controller succeeds in transmitting a packet, this sequence of control commands will be available to the actuator and can be applied without interruption. In the standard implementation, on the other hand, the likelihood that consecutive packet transmissions will be successful is only $(1-p_{\ell})^{2}$. Therefore, the state will often not be brought back close to the origin after an event triggering.

To conform with the guidelines of reproducible research, the {\tt R/Matlab} file to generate the results presented above is publically available at \url{people.kth.se/~demirel/publication}.


\section{Conclusions} \label{sec:Conclusion}

In this paper, we developed a theoretical framework to analyze the trade-off between the communication cost and the control performance of an event-triggering algorithm for control over an unreliable network. We assumed that a threshold-based event-triggering algorithm governs the channel used to transmit the information from the controller to the actuator. Additionally, we assumed the presence of a buffer at the actuator to store the control command sequence received from the controller to mitigate the detrimental effect of packet loss. Furthermore, we developed a multi-dimensional Markov chain model, which characterizes the attempted and successful transmissions of control signals over an unreliable communication link. By combining this communication model with an analytical model of closed-loop performance we provided a systematic way to analyze the trade-off between the communication cost and control performance by appropriately selecting an event threshold.


\section{Appendix}\label{sec:append}


{\it Proof of Lemma~\ref{lem:MarkovChain1}:}
We begin by proving that the process $\{r_{k}\}_{k\in\mathbb{N}_{0}}$ is a Markov chain. Let us consider the case where $x_{k+1}$ belongs to $\mathbb{R}$ which is an uncountable state space. Using the total law of probabilities, we have:
\begin{align*}
	\mathbf{P}\big( r_{k+1} \mid r_{k}, & r_{k-1}, \cdots, r_{0} \big) =
	\int_{\mathbb{R}}\mathbf{P}\big( r_{k+1}, x_{k+1} \mid r_{k}, r_{k-1}, \cdots, r_{0} \big) dx_{k+1} \\
	\stackrel{(a)}{=}&\; \int_{\mathbb{R}}\mathbf{P}\big( r_{k+1} \mid x_{k+1}, r_{k}, r_{k-1}, \cdots, r_{0} \big)\mathbf{P}\big( x_{k+1} \mid r_{k}, r_{k-1}, \cdots, r_{0} \big) dx_{k+1} \\
	\stackrel{(b)}{=}&\; \int_{\mathbb{R}}\mathbf{P}\big( r_{k+1} \mid x_{k+1}, r_{k} \big)\mathbf{P}\big( x_{k+1} \mid r_{k}\ \big) dx_{k+1} \\
	\stackrel{(c)}{=}&\; \int_{\mathbb{R}}\mathbf{P}\big( r_{k+1}, x_{k+1} \mid r_{k} \big) dx_{k+1} \\
	=&\; \mathbf{P}\big( r_{k+1} \mid r_{k}\ \big) 
\end{align*}
where ($a$) and ($c$) come from the definition of the conditional probability, and ($b$) holds because $x_{k+1}$ depends stochastically only on $r_{k}$ as described in~\eqref{eq:IncStep2}, and $r_{k+1}$ depends on $x_{k+1}$ and $r_{k}$ as described in~\eqref{eq:IncStep}. It is also worth noting that knowing $r_{k}=j$ implies that knowing $r_{k}=j, \cdots, r_{k-j}=0$.
Therefore, the process $\{r_{k}\}$ is a Markov chain.

We next proceed to show the ergodicity of this Markov chain.
Since the Markov chain $\{r_{k}\}$, shown in Fig.~\ref{fig:Markov_Chain_wo_Packet_Loss}, has positive transition probabilities for any $\epsilon>0$, the chain is clearly irreducible. Furthermore, the chain is aperiodic because state $\{r_{k}=0\}$ has a non-zero returning loop for any $\epsilon>0$.
By~\cite[Thm~3.3]{Bre:99}, this irreducible chain with finite state space $\mathcal{B}$ is positive recurrent. The process $\{r_{k}\}$ is irreducible, aperiodic and positive recurrent; therefore, it is also ergodic. As $\{r_{k}\}$ is an irreducible aperodic Markov chain with finitely many states, it has a unique invariant distribution $\boldsymbol{\pi}$ such that $\boldsymbol{\pi} P=\boldsymbol{\pi}$ and $\boldsymbol{\pi}\mathbf{1}=1$; see~\cite[Cor~2.11]{Cin:75}. This concludes our proof.
$\hfill\blacksquare$

{\it Proof of Lemma~\ref{lem:Transition_Probabilities}:}
We focus on the case when $i\in\{0,\cdots,T-1\}$, $j=0$ because the other expressions are obvious from the structure of the Markov chain in Fig.~\ref{fig:Markov_Chain_wo_Packet_Loss}. Let us consider the transition probability $p_{00}$. Since $r_{k}=0$ is equivalent to $\phi_{k}=0$, we have:
\begin{align*}
p_{00} = &\; \mathbf{P}\big(r_{k+1}=0 \;\vert\; r_{k}=0\big) \\
		  = &\; \mathbf{P}\big(\vert~w_{k}~\vert>\epsilon \;\vert\; \phi_{k} = 0 \big) \\
		  \stackrel{(a)}{=} &\; \mathbf{P}\big( \vert~w_{k}~\vert>\epsilon \big) = \mathbf{P}\big( \vert~\xi_{0}~\vert>\epsilon \big)\;,
\end{align*}
where ($a$) holds because $\phi_{k}$ is independent of the process noise at time step $k\in\mathbb{N}_{0}$. Similarly, for any $i\in\{1,\cdots,T-1\}$, we have that:
\begin{align*}
p_{i0} = &\; \mathbf{P}\big( r_{k+1}=0~\big\vert~r_{k}=i \big) \\
		  \stackrel{(b)}{=} &\; \mathbf{P}\big( r_{k+1}=0~\big\vert~r_{k}=i, r_{k-1}=i-1, \cdots, r_{k-i}=0 \big) \\
		  \stackrel{(c)}{=} &\; \mathbf{P}\big( \vert~\xi_{i}~\vert>\epsilon~\big\vert~\vert~\xi_{i-1}~\vert\leq\epsilon, \cdots, \vert~\xi_{0}~\vert\leq\epsilon,~\phi_{k-i}=0 \big) \\
		  \stackrel{(d)}{=} &\; \mathbf{P}\big( \vert~\xi_{i}~\vert>\epsilon~\big\vert~\vert~\xi_{i-1}~\vert\leq\epsilon, \cdots, \vert~\xi_{0}~\vert\leq\epsilon \big) \\
		  = &\; \frac{\mathbf{P}\big( \vert~\xi_{i}~\vert>\epsilon, \mathcal{F}_{i} \big)}{\mathbf{P}\big( \mathcal{F}_{i} \big)} = 1 - \frac{\mathbf{P}\big( \mathcal{F}_{i+1} \big)}{\mathbf{P}\big( \mathcal{F}_{i} \big)} \;,
\end{align*}
where ($b$) comes from the Markov property, ($c$) follows the definitions in~\eqref{eq:RandomVarA}, and ($d$) holds because $\phi_{k-i}$ is independent of the process noise after time step $k-i$, and in particular, $\xi_{i}$. Lastly, the combination of these expressions and~\eqref{eq:ComputeProbability} yields the desired result, evaluated by Gaussian integrals~\eqref{eq:GaussianIntegral}.
$\hfill\blacksquare$

{\it Proof of Theorem~\ref{thm:CommunicationRate_Scalar_wo_Losses}:}
The Markov chain $\{r_{k}\}$, depicted in Fig.~\ref{fig:Markov_Chain_wo_Packet_Loss}, is aperiodic, positive recurrent, finite, and irreducible, therefore; the Markov process has a limiting distribution which is also a stationary distribution $(\pi_{i} : i\in\mathcal{B})$. By the ergodic property, i.e., $\pi_{j}=\sum_{j=0}^{\mathrm{T}}\pi_{j}p_{ji}$, we obtain:
\begin{align}
	\pi_{0} = &\; \sum_{j=0}^{\mathrm{T}} \pi_{j} p_{j,0}, \\ 
	\pi_{i} = &\; \big(1-p_{i-1,0}\big)\pi_{i-1}, \quad i\in\{1,\cdots, \mathrm{T}\}.
\end{align}
Combining the balance equation $\sum_{j=0}^{\mathrm{T}}\pi_{j}=1$ and the aforementioned equalities, we thus get: 
\begin{align}
	\pi_{0} = &\; \frac{1}{1 + \sum_{n=1}^{\mathrm{T}}\prod_{m=0}^{n-1}(1-p_{m0})} \;.
\end{align}
This concludes our proof. 
$\hfill\blacksquare$

{\it Proof of Theorem~\ref{thm:ControlPerformanceEval_Scalar_wo_Losses}:}
The Markov chain $\{r_{k}\}_{k\in\mathbb{N}_{0}}^{}$ is aperiodic and positive recurrent on the countable state space $\mathcal{B}=\{ 0,\cdots,\mathrm{T} \}$. Therefore, the process $\{r_{k}\}_{k\in\mathbb{N}_{0}}^{}$ is an ergodic Markov chain with stationary distribution $\pi_{j} = \lim\limits_{k\rightarrow\infty}\mathbf{P}(r_{k}=j)$ for all $j\in\mathcal{B}$. By the ergodic theorem for Markov chains~\cite[pp.~111]{Bre:99}, the linear-quadratic loss~\eqref{eqn:ControlLossFunction} converges to
\begin{align}
	J_{\infty} = \lim_{N\rightarrow\infty}\frac{1}{N}\sum_{k=0}^{N}x_{k}^{\intercal}Q_{r_{k}}x_{k} = \lim_{k\rightarrow\infty}\mathbf{E}_{\pi}\big[ x_{k}^{\intercal}Q_{r_{k}}x_{k}^{} \big]. \label{eqn:ErgodicContCost}
\end{align}
Note that \eqref{eqn:ErgodicContCost} holds since $x_{k+1}$ is a function of $r_{k}$, i.e., $x_{k+1}\triangleq\xi_{k}(r_{k})=\sum_{l=0}^{r_{k}}a^{l}w_{k-l}$. By the law of total expectation, we have:
\begin{align*}
	\mathbf{E}_{\pi}\big[ x_{k}^{\intercal}Q_{r_{k}}x_{k}^{} \big] = \sum_{i\in\mathcal{B}}Q_{i}H_{k,i}, 
\end{align*}
where 
\begin{align*}
	H_{k,i}\triangleq\mathbf{E}\big[x_{k}^{2} \mid r_{k}=i \big]\mathbf{P}(r_{k}=i) \;,
\end{align*}
with $Q_{0}=Q_{x}  + \rho K^{\intercal}Q_{u}K$ and $Q_{i}=Q_{x},~\forall i\in\{ 1,\cdots,\mathrm{T} \}$.

Using the law of total expectations and Bayes' rule, we thus obtain:
%
\begin{align}
	\mathbf{E}\big[ x_{k}^{2} \mid r_{k}=i \big] =&\; \sum_{j\in\mathcal{B}}\mathbf{E}\big[ x_{k}^{2} \mid r_{k}=i, r_{k-1}=j \big] \mathbf{P}(r_{k-1}=j \mid r_{k}=i) \;, \nonumber\\
	&= \sum_{j\in\mathcal{B}}\mathbf{E}\big[ x_{k}^{2} \mid r_{k}=i, r_{k-1}=j \big]\mathbf{P}(r_{k}=i \mid r_{k-1}=j)\frac{\mathbf{P}(r_{k-1}=j)}{\mathbf{P}(r_{k}=i)} \;, \nonumber\\ 
	&= \sum_{j\in\mathcal{B}}p_{ji}\mathbf{E}\big[ x_{k}^{2} \mid r_{k}=i, r_{k-1}=j \big]\frac{\mathbf{P}(r_{k-1}=j)}{\mathbf{P}(r_{k}=i)} \;. 
	\label{eqn:LTE_BayesRule}
\end{align}

By using~\eqref{eqn:LTE_BayesRule}, for $r_{k}=0$, we obtain that:
\begin{small}
\begin{align}
	H_{k,0} &=\; \sum_{j=0}^{\mathrm{T}}p_{j0}\mathbf{E}\big[ x_{k}^{2} \mid r_{k}=0, r_{k-1}=j \big]\mathbf{P}(r_{k-1}=j) \nonumber\\
	&=\; p_{00}\mathbf{E}\big[ x_{k}^{2} \mid \vert x_{k} \vert>\epsilon, \phi_{k-1}=0 \big]\mathbf{P}(r_{k-1}=0) \nonumber\\
	   &\; + \sum_{j=1}^{\mathrm{T}-1}p_{j0}\mathbf{E}\big[ x_{k}^{2} \mid \vert x_{k} \vert>\epsilon, \vert x_{k-n} \vert\leq\epsilon, 1\leq n\leq j, \phi_{k-j-1}=0 \big]\mathbf{P}(r_{k-1}=j) \nonumber\\	   	   
	   &\; + \mathbf{E}\big[ x_{k}^{2} \mid \vert x_{k-n} \vert\leq\epsilon, 1\leq n\leq\mathrm{T}, \phi_{k-\mathrm{T}-1}=0 \big]\mathbf{P}(r_{k-1}=\mathrm{T}) \nonumber\\	   
	&=\; p_{00}\mathbf{E}\big[ x_{k}^{2} \mid \vert x_{k} \vert>\epsilon, \phi_{k-1}=0 \big]\mathbf{P}(r_{k-1}=0) \nonumber\\
	   &\; + \sum_{j=1}^{\mathrm{T}-1}p_{j0}\mathbf{E}\big[ x_{k}^{2} \mid \vert x_{k} \vert>\epsilon, \vert x_{k-n} \vert\leq\epsilon, 1\leq n\leq j, \phi_{k-j-1}=0 \big]\mathbf{P}(r_{k-1}=j) \nonumber\\	   	   
	   &\; + \Big(a^{2}\mathbf{E}\big[ x_{k-1}^{2} \mid \vert x_{k-n} \vert\leq\epsilon, 1\leq n\leq\mathrm{T}, \phi_{k-\mathrm{T}-1}=0 \big] + \sigma_{w}^{2}\Big)\mathbf{P}(r_{k-1}=\mathrm{T}) \;. \label{eqn:performance_eqn1}
\end{align}
\end{small}
Similarly, for any $i\in\{ 1,2,\cdots,\mathrm{T} \}$, we have that:
\begin{small}
\begin{align}
	H_{k,i} =&\; p_{i-1,i}\mathbf{E}\big[ x_{k}^{2} \mid r_{k}=i, r_{k-1}=i-1 \big] \mathbf{P}(r_{k-1}=i-1) \nonumber\\
	=&\; p_{i-1,i}\mathbf{E}\big[x_{k}^{2} \mid \vert x_{k-n} \vert\leq\epsilon, 0\leq n<i, \phi_{k-i}=0 \big]\mathbf{P}(r_{k-1}=i-1) \;. \label{eqn:performance_eqn2} 
\end{align}
\end{small}
Substitution of $x_{k}\triangleq\xi_{k-1}(i)$ into~\eqref{eqn:performance_eqn1} and then into~\eqref{eqn:performance_eqn2} yields
\begin{small}
\begin{align}
	H_{k,0} =&\; p_{00}\mathbf{E}\big[ \xi_{k-1}^{2}(0) \mid \vert \xi_{k-1}(0) \vert>\epsilon \big]\mathbf{P}(r_{k-1}=0) \nonumber\\
	   +&  \sum_{j=1}^{\mathrm{T}-1}p_{j0}\mathbf{E}\big[ \xi_{k-1}^{2}(j) \mid \vert \xi_{k-1}(j) \vert>\epsilon,\vert \xi_{k-n-1}(j-n) \vert\leq\epsilon, 1\leq n\leq j \big]\mathbf{P}(r_{k-1}=j) \nonumber\\	   	   
	   +& \Big(a^{2}\mathbf{E}\big[ \xi_{k-2}^{2}(\mathrm{T}-1) \mid \vert \xi_{k-n-1}(\mathrm{T}-n) \vert\leq\epsilon, 1\leq n\leq\mathrm{T} \big] + \sigma_{w}^{2}\Big)\mathbf{P}(r_{k-1}=\mathrm{T}) \;, \label{eqn:performance_eqn1a} \\ 
	H_{k,i} =&\; p_{i-1,i}\mathbf{E}\big[\xi_{k-1}^{2}(i) \mid \vert \xi_{k-n-1}(i-n) \vert\leq\epsilon, 0\leq n<i \big]\mathbf{P}(r_{k-1}=i-1) \;, \label{eqn:performance_eqn2a}
\end{align}
\end{small}
for all $i\in\{1,\cdots,\mathrm{T}\}$. 
Let's define $H_{i}\triangleq\lim\limits_{k\rightarrow\infty}H_{k,i}$ for all $i\in\mathcal{B}$. Letting $k\rightarrow\infty$, we can rewrite~\eqref{eqn:performance_eqn1a} and~\eqref{eqn:performance_eqn2a} as 
\begin{small}
\begin{align}
	H_{0} =&\; p_{00}\textbf{Var}\big[ \xi_{0} \mid \vert\xi_{0}\vert>\epsilon \big]\pi_{0} + \sum_{j=1}^{\mathrm{T}-1}p_{j0}\textbf{Var}\big[ \xi_{j} \mid \vert\xi_{j}\vert>\epsilon, \mathcal{F}_{j} \big]\pi_{j} + \Big(a^{2}\textbf{Var}\big[ \xi_{\mathrm{T}-1} \mid \mathcal{F}_{\mathrm{T}} \big] + \sigma_{w}^{2}\Big)\pi_{\mathrm{T}} \;, \label{eqn:performance_eqn1b} \\
	H_{i} =&\; p_{i-1,i}\textbf{Var}[\xi_{i-1} \mid \mathcal{F}_{i} \big]\pi_{i-1} = \textbf{Var}[\xi_{i-1} \mid \mathcal{F}_{i} \big]\pi_{i} \;. \label{eqn:performance_eqn2b}
\end{align}
\end{small}
Lastly, substitution of \eqref{eqn:performance_eqn1b},~\eqref{eqn:performance_eqn2b} and $\pi_{i}=\prod_{j=0}^{i-1}(1-p_{j0})\pi_{0}$ into $J_{\infty}=\lim\limits_{k\rightarrow\infty}\mathbf{E}_{\pi}\big[ x_{k}^{\intercal}Q_{r_{k}}x_{k}^{} \big] = \sum_{i\in\mathcal{B}}Q_{i}H_{i}$ establishes~\eqref{eqn:ControlPerformanceEval_wo_loss}. $\hfill\blacksquare$

{\it Proof of Lemma~\ref{lem:MarkovChain2}:}
As in the proof of Lemma~\ref{lem:MarkovChain1}, the similar reasoning applies to show the bidimensional process $(r_{k},\gamma_{k})$ is a Markov chain. Firstly, for $x_{k+1}\in\mathbb{R}$, we write:
\begin{align*}
	\mathbf{P}\big(& r_{k+1},\gamma_{k+1} \mid r_{k}, r_{k-1}, \cdots, r_{0}; \gamma_{k}, \gamma_{k-1}, \cdots, \gamma_{0} \big) \\
	=&\; \int_{\mathbb{R}}	\mathbf{P}\big( r_{k+1},\gamma_{k+1},x_{k+1} \mid r_{k}, \cdots, r_{0}; \gamma_{k}, \cdots, \gamma_{0} \big)dx_{k+1} \\
	\stackrel{(a)}{=}&\; \int_{\mathbb{R}}	\mathbf{P}\big( r_{k+1},\gamma_{k+1} \mid x_{k+1}; r_{k}, \cdots, r_{0}; \gamma_{k}, \cdots, \gamma_{0} \big)\mathbf{P}\big( x_{k+1} \mid r_{k}, \cdots, r_{0}; \gamma_{k}, \cdots, \gamma_{0} \big)dx_{k+1} \\
	\stackrel{(b)}{=}&\; \int_{\mathbb{R}}	\mathbf{P}\big( r_{k+1},\gamma_{k+1} \mid x_{k+1}, r_{k}, \gamma_{k} \big)\mathbf{P}\big( x_{k+1} \mid r_{k}, \gamma_{k}\big)dx_{k+1} \\
	\stackrel{(c)}{=}&\; \int_{\mathbb{R}}	\mathbf{P}\big( r_{k+1},\gamma_{k+1}, x_{k+1} \mid r_{k}, \gamma_{k} \big)dx_{k+1} \\
	=&\; \mathbf{P}\big( r_{k+1},\gamma_{k+1} \mid r_{k}, \gamma_{k} \big)
\end{align*}
where ($a$) and ($c$) come from the definition of the conditional probability, and ($b$) holds because $x_{k+1}$ depends only stochastically on $r_{k}$ and $\gamma_{k}$; $r_{k+1}$ and $\gamma_{k+1}$ depend on $x_{k+1}$, $r_{k}$ and $\gamma_{k}$. 
Notice that knowing $r_{k}=i$ and $\gamma_{k}=j$ implies knowing $r_{k}=i,\cdots, r_{k-i}=0$ and $\gamma_{k}=j,\cdots, \gamma_{k-i}=0$.
This concludes that the process $(r_{k},\gamma_{k})$ is a Markov chain.

The proof is completed by showing the ergodicity of this Markov chain.
As can be seen in Fig.~2, this chain is clearly irreducible. Moreover, the chain is aperiodic since it is irreducible and state $(r_{k},\gamma_{k})=(0,0)$ has a non-zero returning loop for any threshold $\epsilon>0$. The distribution of the return time to state $(0,0)$ is
\begin{align*}
	\mathbf{P}_{0}\big( T_{0} = 1 \big) = (1-p_{\ell})p_{00} \;,
\end{align*}
and, for $n\geq 2$ and $m\geq 0$, 
\begin{align*}
	\mathbf{P}_{0}\big( T_{0} = n + m \big) = (1-p_{\ell})p_{\ell}^{m}(1-p_{00})\cdots (1-p_{n-2,0})p_{n-1,0} \;.
\end{align*}

Using the aforementioned distribution, we can write the expected return times as
\begin{small}
\begin{align*}
	&\mathbf{E}_{0}\big[ T_{0} \big] = \sum_{n=1}^{\mathrm{T}+1}\sum_{m\geq 0}(n+m)\mathbf{P}_{0}\big( T_{0}=n+m \big) \\
	&= \frac{1}{1-p_{\ell}}\underbrace{\Bigg[ p_{00} + \sum_{i=1}^{\mathrm{T}}p_{i0}\prod_{j=0}^{i-1}(1-p_{j0})\Bigg]}_{= 1} 
	+ \sum_{i=1}^{\mathrm{T}+1}i\underbrace{p_{i0}\prod_{j=0}^{i-1}(1-p_{j0})}_{\leq 1} \\
	&\leq \frac{1}{1-p_{\ell}}+\sum_{i=1}^{\mathrm{T}}i = \frac{1}{1-p_{\ell}} + \frac{\mathrm{T}(\mathrm{T}+1)}{2} < \infty
\end{align*}
\end{small}
%
%
and therefore the chain is positive recurrent. 
The process $(r_{k},\gamma_{k})$ is an ergodic Markov chain because it is irreducible, aperiodic and positive recurrent. By [43, Thm~2.1], we conclude that this Markov chain has an invariant distribution $\boldsymbol{\pi}>0$ such that $\pi_{j}=\sum_{i\in\hat{\mathcal{B}}}\pi_{i}p_{ji}$ for all $j\in\hat{\mathcal{B}}$ and $\sum_{j\in\hat{\mathcal{B}}}\pi_{j}=1$. This proves the lemma.
$\hfill\blacksquare$

{\it Proof of Theorem~\ref{thm:CommunicationRate_Scalar_with_Losses}:}
The Markov chain $(r_{k},\gamma_{k}),~k\in\mathbb{N}_{0}$ (shown in Fig.~\ref{fig:MarkovChain_with_Packet_Loss}) is aperiodic, positive recurrent and irreducible, therefore; the Markov process has stationary distributions $(\pi_{ij}^{\infty} : i\in\mathcal{S}, j\in\mathbb{N}_{0} )$. Using the ergodic property, we write that:
\begin{align}
	\pi_{00} =&\; (1-p_{\ell})\sum_{i=0}^{\mathrm{T}}p_{i0}\pi_{i0} + (1-p_{\ell})\sum_{i=0}^{\mathrm{T}}\sum_{j=1}^{\infty}\pi_{ij} \;, \label{eqn:stationary_loss_1} \\
	\pi_{i0} =&\; (1-p_{i-1,0})\pi_{i-1,0} \;, \qquad i\in\{ 1, \cdots, \mathrm{T} \} \;, \label{eqn:stationary_loss_2}
\end{align}
and
\begin{align}
	\pi_{ij} = 
	\begin{cases} 
		p_{\ell}p_{i0}\pi_{i0} &\mbox{if } j = 1,  \\ 
		p_{\ell}\pi_{i,j-1} & \mbox{if } j \neq 1,
	\end{cases} 
	\label{eqn:stationary_loss_3}
\end{align}
for $i\in\{ 0,1, \cdots, \mathrm{T} \}$. By combining~\eqref{eqn:stationary_loss_1}, \eqref{eqn:stationary_loss_2} and \eqref{eqn:stationary_loss_3}, we have that:
\begin{small}
\begin{align}
	\sum_{i=0}^{\mathrm{T}}\sum_{j=0}^{\infty}\pi_{ij} =&\; \sum_{i=0}^{\mathrm{T}}\big( \pi_{i0} + \pi_{i1} + \pi_{20} + \cdots \big)
	=\; \sum_{i=0}^{\mathrm{T}}\pi_{i0} + \frac{1}{1-p_{\ell}}\sum_{i=0}^{\mathrm{T}}\pi_{i1}\;, \nonumber\\
	=&\; \bigg[ 1 + \frac{p_{\ell}}{1-p_{\ell}}p_{00} \bigg]\pi_{00} + \sum_{i=1}^{\mathrm{T}} \bigg[ 1 + \frac{p_{\ell}}{1-p_{\ell}}p_{i0} \bigg]\pi_{i0}\;, \nonumber\\
	=&\; \bigg[ \frac{1-p_{\ell}(1-p_{00})}{1-p_{\ell}} + \sum_{j=1}^{\mathrm{T}} \frac{1-p_{\ell}(1-p_{j0})}{1-p_{\ell}}\prod_{i=0}^{j-1}(1-p_{i0}) \bigg]\pi_{00} \;. \label{eqn:stationary_dist_lossy_netw}
\end{align}
\end{small}
Using the balance equation $\sum_{i=0}^{\mathrm{T}}\sum_{j=0}^{\infty}\pi_{ij}^{\infty} = 1$ and~\eqref{eqn:stationary_dist_lossy_netw}, we obtain~\eqref{eqn:stationary_distribution_w_packet_losses}.

Lastly, to compute the attempted communication rate, we only need to set $p_{\ell}=0$. Hence, we have~\eqref{eqn:stationary_distribution_attempted}.
This concludes our proof.
$\hfill\blacksquare$

{\it Proof of Theorem~\ref{thm:ControlPerformanceEval_Scalar_with_Losses}: }
The proof follows an analysis akin to that in the proof of Theorem~\ref{thm:ControlPerformanceEval_Scalar_wo_Losses}. The process $(r_{k},\gamma_{k}),~k\in\mathbb{N}_{0}$ is an ergodic time-homogeneous Markov Chain (see Fig.~\ref{fig:MarkovChain_with_Packet_Loss}) with stationary distribution $\pi_{ij} = \lim\limits_{k\rightarrow\infty}\mathbf{P}\big( r_{k}=i,\gamma_{k}=j \big)$ for all $i\in\mathcal{B}$ and $j\in\mathbb{N}_{0}$. Using the ergodic theorem for Markov chains~\cite[pp.~111]{Bre:99}, we have that:
\begin{align}
	J_{\infty} = \lim_{N\rightarrow\infty}\frac{1}{N}\sum_{k=0}^{N}x_{k}^{\intercal}Q_{r_{k}}x_{k} = \lim_{k\rightarrow\infty}\mathbf{E}_{\pi}\big[x_{k}^{\intercal}Q_{r_{k}}x_{k}\big]. \label{eqn:ErgodicContCost_Loss}
\end{align}
It is worth noting that \eqref{eqn:ErgodicContCost_Loss} holds because $x_{k}$ depends stochastically on $r_{k-1}$ and $\gamma_{k-1}$, i.e., $x_{k}=a^{\gamma_{k}}\sum\limits_{n=0}^{r_{k}}a^{n}w_{k-\gamma_{k}-n}+\sum\limits_{m=0}^{\gamma_{k}}a^{m}w_{k-m}$.
By the law of total expectation, we have:
\begin{align}
	\mathbf{E}_{\pi}\big[x_{k}^{\intercal}Q_{r_{k}}x_{k}\big] = \sum_{i\in\mathcal{B}}\sum_{j\in\mathbb{N}_{0}} Q_{ij}H_{k,ij} \;,
\end{align}
where
\begin{align}
	H_{k,ij} \triangleq\mathbf{E}\big[ x_{k}^{2} \mid r_{k}=i,\gamma_{k}=j\big]\mathbf{P}\big( r_{k}=i,\gamma_{k}=j \big)
\end{align}
with $Q_{00}=Q_{x}+\rho K^{\intercal}Q_{u}K$ and $Q_{ij}=Q_{x}$ for all $i\in\mathcal{B}$ and $j\in\mathbb{N}_{0}$. 

As in the proof of Theorem~\ref{thm:ControlPerformanceEval_Scalar_wo_Losses}, we aim at defining recurrence equations $H_{k,ij}$ for all $i\in\mathcal{B}$ and $j\in\mathbb{N}_{0}$. For $i=j=0$, the recurrence equation equate to 
\begin{small}
\begin{align}
	H_{k,00} =&\; (1-p_{\ell})\bigg[ p_{00}\mathbf{E}\big[ x_{k}^{2} \mid \vert x_{k} \vert>\epsilon, \phi_{k-1}=0 \big]\mathbf{P}\big(r_{k-1}=0,\gamma_{k-1}=0\big) \nonumber\\
	&+ \sum_{i=1}^{\mathrm{T}-1}p_{i0}\mathbf{E}\big[ x_{k}^{2} \mid \vert x_{k} \vert>\epsilon, \vert x_{k-n} \vert\leq\epsilon, 1\leq n\leq i, \phi_{k-i-1}=0 \big]\mathbf{P}\big( r_{k-1}=i,\gamma_{k-1}=0\big) \nonumber\\
	&+ \mathbf{E}\big[ x_{k}^{2} \mid \vert x_{k-n} \vert\leq\epsilon, 1\leq n\leq\mathrm{T} =1, \phi_{k-\mathrm{T}-1}=0 \big]\mathbf{P}\big( r_{k-1}=\mathrm{T},\gamma_{k-1}=0 \big) \nonumber\\
	& + \sum_{j=1}^{\infty}\mathbf{E}\big[ x_{k}^{2} \mid \vert x_{k-j} \vert>\epsilon, \phi_{k-j-1}=0 \big]\mathbf{P}\big( r_{k-1}=0,\gamma_{k-1}=j \big) \nonumber\\
	&+ \sum_{i=1}^{\mathrm{T}-1}\sum_{j=1}^{\infty}\mathbf{E}\big[ x_{k}^{2} \mid \vert x_{k-j} \vert>\epsilon, \vert x_{k-j-n} \vert\leq\epsilon, 1\leq n\leq i, \phi_{k-j-i-1}=0 \big]\mathbf{P}\big( r_{k-1}=i,\gamma_{k-1}=j \big) \nonumber\\
	&+ \sum_{j=1}^{\infty}\mathbf{E}\big[ x_{k}^{2} \mid \vert x_{k-j-n} \vert\leq\epsilon, 1\leq n\leq\mathrm{T}, \phi_{k-j-\mathrm{T}-1}=0 \big]\mathbf{P}\big( r_{k-1}=\mathrm{T},\gamma_{k-1}=j \big) \bigg]\;. \label{eqn:performance_loss_eqn1}
\end{align}
\end{small}
For any $i\in\{ 1,2,\cdots,\mathrm{T} \}$ and $j=0$, we have that:
\begin{small}
\begin{align}
	H_{k,i0}=\; p_{i-1,i}\mathbf{E}\big[ x_{k}^{2} \mid \vert x_{k-n} \vert\leq\epsilon, 0\leq n<i, \phi_{k-i}=0\big]\mathbf{P}\big( r_{k-1}=i-1,\gamma_{k-1}=0 \big) \;. \label{eqn:performance_loss_eqn2}
\end{align}
\end{small}
For any $i\in\{0,1,\cdots,\mathrm{T}\}$ and $j=1$, we have that:
\begin{small}
\begin{align}
	H_{k,i1} =\; p_{\ell}p_{i0}\mathbf{E}\big[ x_{k}^{2} \mid \vert x_{k} \vert>\epsilon, \vert x_{k-n} \vert\leq\epsilon, 1\leq n\leq i, 
	\phi_{k-i-1}=0 \big]\mathbf{P}\big( r_{k-1}=i,\gamma_{k-1}=0 \big) \;. \label{eqn:performance_loss_eqn3}
\end{align}
\end{small}
For all $i\in\{0,1,\cdots,\mathrm{T}\}$ and $j\geq 2$, we have that:
\begin{small}
\begin{align}
	H_{k,ij} =\; p_{\ell}\mathbf{E}\big[ x_{k}^{2} \mid \vert x_{k-j+1} \vert>\epsilon, \vert x_{k-n} \vert\leq\epsilon, j\leq n<i+j,
	\phi_{k-i-j}=0 \big]\mathbf{P}\big( r_{k-1}=i,\gamma_{k-1}=j-1 \big) \;. \label{eqn:performance_loss_eqn4}
\end{align}
\end{small}

Substituting $x_{k}\triangleq a^{j}\xi_{k-\gamma_{k}}(i)+\sum_{m=0}^{j}a^{m}w_{k-m}$ into~\eqref{eqn:performance_loss_eqn1}, \eqref{eqn:performance_loss_eqn2}-\eqref{eqn:performance_loss_eqn4} yields 
\begin{small}
\begin{align}
	H_{k,00} =&\; (1-p_{\ell})\bigg[p_{00}\mathbf{E}\big[ \xi_{k-1}^{2}(0) \mid \vert\xi_{k-1}(0)\vert>\epsilon \big]\mathbf{P}\big(r_{k-1}=0,\gamma_{k-1}=0\big) \nonumber\\
	&+ \sum_{i=1}^{\mathrm{T}-1}p_{i0}\mathbf{E}\big[ \xi_{k-1}^{2}(i) \mid \vert\xi_{k-1}(i)\vert>\epsilon, \vert\xi_{k-n-1}(i-n)\vert\leq\epsilon, 1\leq n\leq i \big]\mathbf{P}\big( r_{k-1}=i,\gamma_{k-1}=0 \big) \nonumber\\
	&+ \Big( a^{2}\mathbf{E}\big[ \xi_{k-2}^{2}(\mathrm{T}-1) \mid \vert\xi_{k-n-1}(\mathrm{T}-n)\vert\leq\epsilon, 1\leq n\leq\mathrm{T} \big] + \sigma_{w}^{2}\Big)\mathbf{P}\big( r_{k-1}=\mathrm{T},\gamma_{k-1}=0 \big) \nonumber\\
	&+ \sum_{j=1}^{\infty}\bigg( a^{2j}\mathbf{E}\big[ \xi_{k-j-1}^{2}(0) \mid \vert\xi_{k-j-1}(0)\vert>\epsilon \big] + \sigma_{w}^{2}\sum_{l=0}^{j-1}a^{2l} \bigg)\mathbf{P}\big(r_{k-1}=0,\gamma_{k-1}=j\big) \nonumber\\
	&+ \sum_{i=1}^{\mathrm{T}-1}\sum_{j=1}^{\infty}\bigg(a^{2j}\mathbf{E}\big[ \xi_{k-j-1}^{2}(i) \mid \vert\xi_{k-j-1}(i)\vert>\epsilon, \vert\xi_{k-j-n-1}(i-n)\vert\leq\epsilon, 1\leq n\leq i \big] + \sigma_{w}^{2}\sum_{l=0}^{j-1}a^{2l} \bigg) \nonumber\\
	&\times\mathbf{P}\big( r_{k-1}=i,\gamma_{k-1}=j \big) \nonumber\\
	&+ \sum_{j=1}^{\infty}\bigg(a^{2j}\Big( a^{2}\mathbf{E}\big[ \xi_{k-j-2}^{2}(\mathrm{T}-1) \mid \vert\xi_{k-j-n-1}(\mathrm{T}-n)\vert\leq\epsilon, 1\leq n\leq\mathrm{T} \big] + \sigma_{w}^{2}\Big) + \sigma_{w}^{2}\sum_{l=0}^{j-1}a^{2l} \bigg)\nonumber\\
	&\times\mathbf{P}\big( r_{k-1}=\mathrm{T},\gamma_{k-1}=j \big) \bigg] \label{eqn:performance_loss_eqn1_a}
\end{align}
\end{small}
and 
\begin{small}
\begin{align}
	H_{k,i0} &=\; p_{i-1,i}\mathbf{E}\big[ \xi_{k-1}^{2}(i-1) \mid \vert\x_{k-n-1}(i-n-1)\vert\leq\epsilon, 0\leq n<i\big]\mathbf{P}\big( r_{k-1}=i-1,\gamma_{k-1}=0 \big), \label{eqn:performance_loss_eqn2_a}\\
	H_{k,i1} &=\; p_{\ell}p_{i0}\mathbf{E}\big[ \xi_{k-1}^{2}(i) \mid \vert\xi_{k-1}(i)\vert>\epsilon, \vert\xi_{k-n-1}(i-n)\vert\leq\epsilon, 1\leq n\leq i \big]\mathbf{P}\big( r_{k-1}=i,\gamma_{k-1}=0 \big), \label{eqn:performance_loss_eqn3_a} \\
	H_{k,ij} &=\; p_{\ell}\bigg[ a^{2(j-1)}\mathbf{E}\big[ \xi_{k-1}^{2}(i) \mid \vert\xi_{k-1}(i)\vert>\epsilon, \vert\xi_{k-n-1}(i-n)\vert\leq\epsilon, 1\leq n\leq i \big] + \sigma_{w}^{2}\sum_{l=0}^{j-2}a^{2l} \bigg] \nonumber\\
	&\times\mathbf{P}\big( r_{k-1}=i,\gamma_{k-1}=j-1 \big). \label{eqn:performance_loss_eqn4_a}
\end{align}
\end{small}

We next define $H_{ij}\triangleq\lim_{k\rightarrow\infty}H_{k,ij}$ for all $i\in\mathcal{B}$ and $j\in\mathbb{N}_{0}$. Letting $k\rightarrow\infty$, we can rewrite~\eqref{eqn:performance_loss_eqn1_a} as 
\begin{small}
\begin{align}
	H_{00}=&\; (1-p_{\ell})\Bigg[ \sum_{i=0}^{\mathrm{T-1}}p_{i0}\textbf{Var}\big[ \delta_{i} \mid \vert\delta_{i}\vert>\epsilon,\mathcal{F}_{i} \big]\pi_{i0} + p_{\mathrm{T}0}\Big(a^{2}\textbf{Var}\big[ \delta_{\mathrm{T}-1} \mid \mathcal{F}_{\mathrm{T}} \big] + \sigma_{w}^{2}\Big)\pi_{\mathrm{T}0} \nonumber\\
	& + \sum_{i=0}^{\mathrm{T}-1}\sum_{j=1}^{\infty}\bigg[ a^{2j}\textbf{Var}\big[ \delta_{i} \mid\vert\delta_{i}\vert>\epsilon,\mathcal{F}_{i} \big] + \sigma_{w}^{2}\sum_{l=0}^{j-1}a^{2l} \bigg]\pi_{ij} + \sum_{j=1}^{\infty}\bigg[ a^{2j}\Big( a^{2}\textbf{Var}\big[ \delta_{\mathrm{T}-1} \mid \mathcal{F}_{\mathrm{T}} \big] + \sigma_{w}^{2} \Big) \nonumber\\
	& + \sigma_{w}^{2}\sum_{l=0}^{j-1}a^{2l} \bigg]\pi_{\mathrm{T}j} \Bigg] \;. \label{eqn:performance_loss_eqn1_b}
\end{align}
\end{small}

Note that $H_{00}$ contains the summations of infinite geometric series, and hence $H_{00}$ converges to an fixed value if the condition $p_{\ell}a^{2} < 1$ holds.
Thereby, with the notation $\Sigma_{i}^{c}\triangleq\mathbf{Var}\big[ \delta_{i} \mid \vert\delta_{i}\vert > \epsilon, \mathcal{F}_{i} \big], i\in\{0,\cdots,\mathrm{T}-1\}$ and $\Sigma_{\mathrm{T}}^{c}\triangleq a^{2}\mathbf{Var}\big[ \delta_{\mathrm{T}-1} \mid \mathcal{F}_{\mathrm{T}} \big] + \sigma_{w}^{2}$, we rewrite~\eqref{eqn:performance_loss_eqn1_b} as
\begin{small}
\begin{align}
	H_{00} =&\; (1-p_{\ell})\sum_{i=0}^{\mathrm{T}}p_{i0}\Sigma_{i}^{c}\pi_{i0} + (1-p_{\ell})\sum_{i=0}^{\mathrm{T}}\Bigg[ a^{2}\Sigma_{i}^{c}\sum_{j=0}(p_{\ell}a^{2})^{j} + \sigma_{w}^{2}\sum_{j=0}p_{\ell}^{j}\sum_{l=0}^{j}a^{2l} \Bigg]\pi_{i1} \nonumber\\
	%
	%
	=&\; (1-p_{\ell})\sum_{i=0}^{\mathrm{T}}p_{i0}\Bigg[ \frac{p_{\ell}a^{2}\Sigma_{i}^{c}}{1-p_{\ell}a^{2}} + \Sigma_{i}^{c} + \frac{p_{\ell}\sigma_{w}^{2}}{1-a^{2}}\bigg( \sum_{j=0}p_{\ell}^{j} - a^{2}\sum_{j=0}(p_{\ell}a^{2})^{j} \bigg) \Bigg]\pi_{i0} \nonumber\\
	=&\; (1-p_{\ell})\sum_{i=0}^{\mathrm{T}}p_{i0}\Bigg[ \frac{\Sigma_{i}^{c}}{1-p_{\ell}a^{2}}  + \frac{p_{\ell}\sigma_{w}^{2}}{(1-p_{\ell})(1-p_{\ell}a^{2})} \Bigg]\pi_{i0} \nonumber\\
	=&\; \frac{1-p_{\ell}}{1-p_{\ell}a^{2}}\sum_{i=0}^{\mathrm{T}}p_{i0}\bigg[ \Sigma_{i}^{c} +\frac{p_{\ell}\sigma_{w}^{2}}{1-p_{\ell}}  \bigg]\pi_{i0} \;. \label{eqn:performance_loss_eqn1_c}
\end{align}
\end{small}

As $k\rightarrow\infty$,~\eqref{eqn:performance_loss_eqn2_a}--\eqref{eqn:performance_loss_eqn4_a} become
\begin{small}
\begin{align}
	H_{i0}=&\; p_{i-1,i}\textbf{Var}\big[\delta_{i-1} \mid N_{i} \big]\pi_{i-1,0} = \Sigma_{i-1}\pi_{i0}, \label{eqn:performance_loss_eqn2_b}\\
	H_{i1}=&\; p_{\ell}p_{i0}\textbf{Var}\big[ \delta_{i} \mid \vert\delta_{i}\vert>\epsilon, N_{i} \big]\pi_{i0}^{\infty} = \Sigma_{i}^{c}\pi_{i1}, \label{eqn:performance_loss_eqn3_b}\\
	H_{ij} =&\; p_{\ell}\bigg[ a^{2(j-1)}\textbf{Var}\big[ \delta_{i} \mid \vert\delta_{i}\vert>\epsilon, N_{i} \big] + \sum_{l=0}^{j-2}a^{2l}\sigma_{w}^{2} \bigg]\pi_{i,j-1} =\; \bigg[ a^{2(j-1)}\Sigma_{i}^{c} + \sum_{l=0}^{j-2}a^{2l}\sigma_{w}^{2} \bigg]\pi_{ij}. \label{eqn:performance_loss_eqn4_b}
\end{align}
\end{small}

By keeping $\lim\limits_{k\rightarrow\infty}\mathbf{E}\big[ x_{k}^{\intercal}Q_{r_{k}}x_{k}\big]=\sum_{i\in\mathcal{B}}\sum_{j\in\mathbb{N}_{0}}Q_{ij}H_{ij}$ in mind, we seperate the control loss function into two parts: $J_{\infty} = J_{\mathrm{ctrl}} + J_{\mathrm{loss}}$. Firstly, we deal with $J_{\mathrm{ctrl}}$ as follows:
\begin{small}
\begin{align}
	J_{\mathrm{ctrl}}  =\; Q_{00}C_{00} + Q_{x}\sum_{i=1}^{\mathrm{T}}H_{i0} =\; \frac{(1-p_{\ell})Q_{00}}{1-p_{\ell}a^{2}}\sum_{i=0}^{\mathrm{T}}p_{i0}\bigg[ \Sigma_{i}^{c} +\frac{p_{\ell}\sigma_{w}^{2}}{1-p_{\ell}} \bigg]\pi_{i0} + Q_{x}\sum_{i=1}^{\mathrm{T}}\Sigma_{i-1}\pi_{i0} \;.
	\label{eqn:ControlLoss_LossyNetworks_Part1}
\end{align}
\end{small}
Next, we obtain the performance deterioration $J_{\mathrm{loss}}$, caused by packet losses, in terms of control loss as follows:
\begin{small}
\begin{align}
	J_{\mathrm{loss}} =&\; Q_{x}\sum_{i=0}^{\mathrm{T}}\sum_{j=1}^{\infty}\mathbf{E}\big[ x_{k}^{2} \mid r_{k}=i,\gamma_{k}=j \big]\pi_{ij} \nonumber\\
	=&\; Q_{x}\sum_{i=0}^{\mathrm{T}}\Bigg[\Sigma_{i}^{c}\pi_{i1} + \sum_{j=2}^{\infty}\bigg[ a^{2(j-1)}\Sigma_{i}^{c} + \sum_{l=0}^{j-2}a^{2l}\sigma_{w}^{2} \bigg]\pi_{ij} \Bigg] \nonumber\\
	%
	=&\; Q_{x}\sum_{i=0}^{\mathrm{T}}\Bigg[ \Sigma_{i}^{c}\sum_{j=0}^{\infty}(p_{\ell}a^{2})^{j} + p_{\ell}\sigma_{w}^{2}\sum_{j=0}^{\infty}p_{\ell}^{j}\sum_{l=0}^{j}a^{2l} \Bigg]\pi_{i1} \nonumber\\
	=&\; Q_{x}\sum_{i=0}^{\mathrm{T}}\Bigg[ \frac{\Sigma_{i}^{c}}{1-p_{\ell}a^{2}} + p_{\ell}\sigma_{w}^{2}\sum_{j=0}^{\infty}p_{\ell}^{j}\frac{1-a^{2(j+1)}}{1-a^{2}} \Bigg]\pi_{i1} \nonumber\\
	%
	%
	=&\; \frac{p_{\ell}Q_{x}}{1-p_{\ell}a^{2}}\sum_{i=0}^{\mathrm{T}}p_{i0}\bigg[\Sigma_{i}^{c}+\frac{p_{\ell}\sigma_{w}^{2}}{1-p_{\ell}}\bigg]\pi_{i0} \;.
	\label{eqn:ControlLoss_LossyNetworks_Part2}
\end{align}
\end{small}

Combining~\eqref{eqn:ControlLoss_LossyNetworks_Part1} and~\eqref{eqn:ControlLoss_LossyNetworks_Part2} with $\pi_{i0}=\prod_{j=0}^{i-1}(1-p_{j0})\pi_{00}$, it is straightforward to varify the control loss function~\eqref{eqn:ControlLoss_LossyNetworks}. This concludes the proof.
$\hfill\blacksquare$

{\it Proof of Lemma~\ref{lem:MarkovChain3}: }
Similar to the one of Lemma~\ref{lem:MarkovChain2}.
$\hfill\blacksquare$

{\it Proof of Lemma~\ref{lem:Transition_Probabilities_Vector}: }
The proof of Lemma~\ref{lem:Transition_Probabilities_Vector} follows similar lines to that of Lemma~\ref{lem:Transition_Probabilities}, but an outline of the proof is included for completeness.
We firstly concentrate on the cases when $i_{1}=i_{0}=0$ $j_{1}=1$, $j_{0}=\nu$, and $i_{1}=i_{0}\in\{ 1,\cdots,\mathrm{T}-1 \}$, $j_{1}=1$, $j_{0}=0$. Similar to~Lemma~\ref{lem:Transition_Probabilities}, the other expressions are obvious from the structure of the Markov chain, depicted in Fig.~\ref{fig:MarkovChain_VectorCase}. We let consider the transition probability $p_{00}$. Since $(r_{k},\eta_{k})=(0,\nu)$ is equivalent to $\bphi_{k}=\mathbf{0}_{n}$, we have:
\begin{align*}
p_{00} = &\; \mathbf{P}\big( (0,1) \mid (0,\nu) \big)\;, \\
		  = &\; \mathbf{P}\big( \parallel \w_{k}\parallel_{\infty}>\epsilon \mid \bphi_{k} = \mathbf{0}_{n} \big)\;, \\
		  \stackrel{(a)}{=} &\; \mathbf{P}\big( \parallel \w_{k}\parallel_{\infty}>\epsilon \big) = \mathbf{P}\big( \parallel\bdel_{0}\parallel_{\infty}>\epsilon \big)\;,
\end{align*}
where $(a)$, as Lemma~\ref{lem:Transition_Probabilities}, holds because $\bphi_{k}$ is independent of the process noise at time step $k\in\mathbb{N}_{0}$. On the other hand, for any $i\in\{1,\cdots,\mathrm{T}-1\}$, we have: 
\begin{align*}
p_{ii} = &\; \mathbf{P}\big( (i,1) \mid (i,0) \big)\;, \\
		  = &\; \mathbf{P}\big( (i,1) \mid (i,0), (i-1,0), \cdots, (1,0), (0,\nu) \big)\;, \\
		  = &\; \frac{\mathbf{P}\big( \parallel\bdel_{i}\parallel_{\infty}>\epsilon, \mathcal{F}_{i} \big)}{\mathbf{P}\big( \mathcal{F}_{i} \big)} = 1 - \frac{\mathbf{P}\big( \mathcal{F}_{i+1} \big)}{\mathbf{P}\big( \mathcal{F}_{i} \big)} \;,
\end{align*}
which follows the same arguments as those of Lemma~\ref{lem:Transition_Probabilities}.
$\hfill\blacksquare$

{\it Proof of Theorem~\ref{thm:CommunicationRate_Vector}: }
Similar to the one of Theorem~\ref{thm:CommunicationRate_Scalar_wo_Losses}.
%
$\hfill\blacksquare$

{\it Proof of Theorem~\ref{thm:ControlPerformanceEval}: }
Similar to the one of Theorem~\ref{thm:ControlPerformanceEval_Scalar_wo_Losses}.
%
$\hfill\blacksquare$

{\it Proof of Theorem~\ref{thm:CommRate_Vector_Loss}: }
Similar to the one of Theorem~\ref{thm:CommunicationRate_Scalar_with_Losses}.
%
$\hfill\blacksquare$

{\it Proof of Theorem~\ref{thm:CntrlPerf_Vector_Loss}: }
Similar to the one of Theorem~\ref{thm:ControlPerformanceEval_Scalar_with_Losses}.
%
$\hfill\blacksquare$

\bibliographystyle{IEEEtran}
\bibliography{ieeetac2013}

\begin{thebibliography}{10}
\providecommand{\url}[1]{#1}
\csname url@samestyle\endcsname
\providecommand{\newblock}{\relax}
\providecommand{\bibinfo}[2]{#2}
\providecommand{\BIBentrySTDinterwordspacing}{\spaceskip=0pt\relax}
\providecommand{\BIBentryALTinterwordstretchfactor}{4}
\providecommand{\BIBentryALTinterwordspacing}{\spaceskip=\fontdimen2\font plus
\BIBentryALTinterwordstretchfactor\fontdimen3\font minus
  \fontdimen4\font\relax}
\providecommand{\BIBforeignlanguage}[2]{{%
\expandafter\ifx\csname l@#1\endcsname\relax
\typeout{** WARNING: IEEEtran.bst: No hyphenation pattern has been}%
\typeout{** loaded for the language `#1'. Using the pattern for}%
\typeout{** the default language instead.}%
\else
\language=\csname l@#1\endcsname
\fi
#2}}
\providecommand{\BIBdecl}{\relax}
\BIBdecl

\bibitem{DGJ:13}
B.~Demirel, V.~Gupta, and M.~Johansson, ``On the trade-off between control
  performance and communication cost for event-triggered control over lossy
  networks,'' in \emph{the $12^{th}$ European Control Conference}, Zurich,
  Switzerland, July, 17-19 2013.

\bibitem{AsW:97}
K.~J. Astr\"{o}m and B.~Wittenmark, \emph{Computer Controlled Systems},
  $3^{rd}$ edition~ed., ser. Information and System Sciences Series.\hskip 1em
  plus 0.5em minus 0.4em\relax Prentice Hall, 1997.

\bibitem{ChF:95}
T.~Chen and B.~Francis, \emph{Optimal Sampled-Data Control Systems}.\hskip 1em
  plus 0.5em minus 0.4em\relax Berlin: Springer Verlag, 1995.

\bibitem{HJT:12}
W.~Heemels, K.~H. Johansson, and P.~Tabuada, ``An introduction to
  event-triggered and self-triggered control,'' in \emph{the IEEE $51^{st}$
  Annual Conference on Decision and Control (CDC'12)}, Maui, HI, 10-13 Dec.
  2012, pp. 3270 -- 3285.

\bibitem{HvK:02}
D.~Hristu-{V}arsakelis and P.~R. Kumar, ``Interrupt-based feedback control over
  a shared communication medium,'' in \emph{$41^{st}$ IEEE Conference on
  Decision and Control (CDC'02)}, 2002.

\bibitem{RSJ:11}
C.~Ramesh, H.~Sandberg, and K.~H. Johansson, ``Steady state performance
  analysis of multiple state-based schedulers with {CSMA},'' in \emph{the
  Proceedings of the $50^{th}$ IEEE Conference on Decision and Control and
  European Control Conference (CDC-ECC'11)}, Orlando, FL, USA, Dec. 12-15 2011.

\bibitem{RSJ:12}
------, ``Stability analysis of multiple state-based schedulers with {CSMA},''
  in \emph{the Proceedings of the $51^{st}$ IEEE Conference on Decision and
  Control (CDC'12)}, Maui, USA, Dec. 2012.

\bibitem{RSJ:13}
------, ``Design of state-based schedulers for a network of control loops,''
  \emph{IEEE Transactions on Automatic Control}, vol.~58, no.~8, pp.
  1962--1975, Aug. 2013.

\bibitem{AMA+:14}
J.~Araujo, M.~M. Jr., A.~Anta, P.~Tabuada, and K.~H. Johansson, ``System
  architectures, protocols and algorithms for aperiodic wireless control
  systems,'' \emph{IEEE Transactions of Industrial Electronics}, vol.~10,
  no.~1, pp. 175 -- 184, Feb. 2014.

\bibitem{DFJ:12}
D.~Dimarogonas, E.~Frazzoli, and K.~Johansson, ``Distributed event-triggered
  control for multi-agent systems,'' \emph{IEEE Transactions on Automatic
  Control}, vol.~57, no.~5, pp. 1291--1297, May 2012.

\bibitem{Lem:10}
M.~Lemmon, \emph{Event-Triggered Feedback in Control, Estimation, and
  Optimization in Networked Control Systems}.\hskip 1em plus 0.5em minus
  0.4em\relax Springer Verlag, vol. 405, ch. Lecture Notes in Control and
  Information Sciences.

\bibitem{Tab:07}
P.~Tabuada, ``Event-triggered real-time scheduling of stabilizing control
  tasks,'' \emph{IEEE Transactions on Automatic Control}, vol.~52, no.~9, pp.
  1680--1685, Sept. 2007.

\bibitem{HDT:13}
W.~P. M.~H. Heemels, M.~C.~F. Donkers, and A.~R. Teel, ``Periodic
  event-triggered control for linear systems,'' \emph{IEEE Transactions on
  Automatic Control}, vol.~58, no.~4, pp. 847--861, Apr. 2013.

\bibitem{AnT:10}
A.~Anta and P.~Tabuada, ``On the minimum attention and anytime attention
  problems for nonlinear systems,'' in \emph{$49^{th}$ IEEE Conference on
  Decision and Control (CDC'10)}, Dec. 2010, pp. 3234--3239.

\bibitem{VMB:09}
M.~Velsaco, P.~Marti, and E.~Bini, ``On {L}yapunov sampling for event-driven
  controllers,'' in \emph{$48^{th}$ IEEE Conference on Decision and
  Control/$28^{th}$ Chinese Control Conference (CDC/CCC'09)}, 2009.

\bibitem{LuL:10}
D.~Lunze and D.~Lehmann, ``A state--feedback approach to event--based
  control,'' \emph{Automatica}, vol.~46, no.~1, pp. 211--215, Nov. 2010.

\bibitem{HeD:13}
W.~P. M.~H. Heemels and M.~C.~F. Donkers, ``Model-based periodic
  event-triggered control for linear systems,'' \emph{Automatica}, vol.~49,
  no.~3, pp. 698--711, March 2013.

\bibitem{MoA:04}
L.~A. Montestruque and P.~Antsaklis, ``Stability of model-based networked
  control systems with time-varying transmission times,'' \emph{IEEE
  Transactions on Automatic Control}, vol.~49, no.~9, pp. 1562--1572, Sept.
  2004.

\bibitem{OMT:02}
P.~Otanez, J.~Moyne, and D.~Tilbury, ``Using deadbands to reduce communication
  in networked control systems,'' in \emph{American Control Conference
  (ACC'02)}, 2002, pp. 3015--3020.

\bibitem{HSB:08}
W.~Heemels, J.~Sandee, and P.~V.~D. Bosch, ``Analysis of event-driven
  controllers for linear systems,'' \emph{International Journal of Control},
  vol.~81, no.~4, pp. 571--590, Apr. 2008.

\bibitem{XGA:12}
M.~Xia, V.~Gupta, and P.~Antsaklis, ``Networked state estimation over a shared
  communication medium,'' \emph{American Control Conference}, pp. 4128--4133,
  17-19 June 2013.

\bibitem{AsB:99}
K.~J. Astr\"{o}m and B.~Bernhardsson, ``Comparison of periodic and event-based
  sampling for first order stochastic systems,'' in \emph{Proceedings of the
  $14^{th}$ IFAC World Congress}, 1999, pp. 301 -- 306.

\bibitem{AsB:02}
------, ``Comparison of {R}iemann and {L}ebesgue sampling for first order
  stochastic systems,'' in \emph{$41^{st}$ IEEE Conference on Decision and
  Control (CDC'02)}, vol.~2, Dec. 2002, pp. 2011--2016.

\bibitem{HJC:08}
T.~Henningsson, E.~Johansson, and A.~Cervin, ``Sporadic event-based control of
  first-order linear stochastic systems,'' \emph{Automatica}, vol.~44, no.~11,
  pp. 2890--2895, Nov. 2008.

\bibitem{Rab:06}
M.~Rabi, ``Packet based inference and control,'' available at
  http://hdl.handle.net/1903/3970, University of Maryland, 2006.

\bibitem{MeC:12}
X.~Meng and T.~Chen, ``Optimal sampling and performance comparison of periodic
  and event based impulse control,'' \emph{IEEE Transactions on Automatic
  Control}, vol.~57, no.~12, pp. 3252--3259, Dec. 2012.

\bibitem{BlA:12}
R.~Blind and F.~Allg\"{o}ver, ``The performance of event-based control the
  performance of event-based control for scalar systems with packet losses,''
  in \emph{Proceedings of the $52^{nd}$ IEEE Conference of Decision and
  Control}, 2012.

\bibitem{RaJ:09}
M.~Rabi and K.~H. Johansson, ``Scheduling packets for event-triggered
  control,'' in \emph{$10^{th}$ European Control Conference (ECC'09)}, Aug.
  2009.

\bibitem{QGM+:14}
D.~E. Quevedo, V.~Gupta, W.~J. Ma, and S.~Yuksel, ``Stochastic stability of
  event-triggered anytime control,'' \emph{IEEE Transactions on Automatic
  Control, To appear}, 2014.

\bibitem{QuN:11}
D.~E. Quevedo and D.~Nesic, ``Input-to-state stability of packetized predictive
  control over unreliable networks affected by packet-dropouts,'' \emph{IEEE
  Transactions on Automatic Control}, vol.~56, no.~2, pp. 370--375, Feb. 2011.

\bibitem{QuN:12}
------, ``Robust stability of packetized predictive control of nonlinear
  systems with disturbances and markovian packet losses,'' \emph{Automatica},
  vol.~48, no.~8, pp. 1803--1811, Aug. 2012.

\bibitem{MaG:12}
W.-J. Ma and V.~Gupta, ``Input-to-state stability of hybrid systems with
  receding horizon control in the presence of packet dropouts,''
  \emph{Automatica}, vol.~48, no.~8, pp. 1920--1923, Aug. 2012.

\bibitem{ZLR:10}
Y.-B. Zhao, G.-P. Liu, and D.~Rees, ``Packet-based deadband control for
  internet-based networked control systems,'' \emph{IEEE Transactions on
  Control Systems Technology}, vol.~18, no.~5, pp. 1057--1067, Sept. 2010.

\bibitem{DFH:13}
M.~Dolgov, J.~Fischer, and U.~D. Hanebeck, ``Event-based {LQG} control over
  networks with random transmission delays and packet losses,'' in \emph{the
  $4^{th}$ IFAC Workshop on Distributed Estimation and Control in Networked
  Systems}, 2013, pp. 23--30.

\bibitem{Rob:95}
C.~P. Robert, ``Simulation of truncated normal variables,'' \emph{Statistics
  and Computing}, vol.~5, pp. 121--125, 1995.

\bibitem{MaW:09}
B.~G. Manjunath and S.~Wilhelm, ``Moments calculation for the double truncated
  multivariate normal density,'' Sept. 11, 2009, Available at SSRN:
  http://ssrn.com/abstract=1472153 or http://dx.doi.org/10.2139/ssrn.1472153.

\bibitem{Ore:81}
J.~O'{R}eilly, ``The discrete linear time invariant time-optimal control
  problem -- {A}n overview,'' \emph{Automatica}, vol.~17, no.~2, pp. 363--370,
  1981.

\bibitem{Hes:09}
J.~Hespanha, \emph{Linear Systems Theory}.\hskip 1em plus 0.5em minus
  0.4em\relax Princeton, New Jersey: Princeton University Press, 2009.

\bibitem{Sch:09}
L.~Schenato, ``To zero or to hold control inputs with lossy links?'' \emph{IEEE
  Transactions on Automatic Control}, vol.~54, no.~5, pp. 1093--1099, May 2009.

\bibitem{CFM:05}
O.~Costa, M.~Fragoso, and R.~Marques, \emph{Discrete-Time Markov Jump Linear
  Systems}, ser. Probability and its Applications.\hskip 1em plus 0.5em minus
  0.4em\relax Springer Verlag, 2005.

\bibitem{ZYL:13}
R.~Zurkowski, S.~Y\"{u}ksel, and T.~Linder, ``On rates of convergence for
  {M}arkov chains under random time state dependent drift criteria,''
  \emph{arXiv:1312.4210v1}, 2013.

\bibitem{FrB:02}
M.~D. Fragoso and J.~Baczynski, ``Lyapunov coupled equations for
  continuous-time infinite markov jump linear systems,'' \emph{Journal of
  Mathematical Analysis and Applications}, vol. 274, no.~1, pp. 319--335, Oct.
  2002.

\bibitem{FrB:06}
------, ``Optimal control for continuous-time linear quadratic problems with
  infinite markov jump parameters,'' \emph{SIAM Journal on Control and
  Optimization}, vol.~40, no.~1, pp. 270--297, July 2006.

\bibitem{ZBP:01}
W.~Zhang, M.~S. Branicky, and S.~M. Phillips, ``Stability of networked control
  systems,'' \emph{IEEE Control Systems}, vol.~21, no.~1, pp. 84 -- 99, Feb.
  2001.

\bibitem{Bre:99}
P.~Bremaud, \emph{Markov Chains: Gibbs fields, Monte Carlo Simulation and
  Queues}, ser. Texts in Applied Mathematics, Vol. 31.\hskip 1em plus 0.5em
  minus 0.4em\relax Springer Verlag, 1999.

\bibitem{Cin:75}
E.~Cinlar, \emph{Introduction to stochastic processes}.\hskip 1em plus 0.5em
  minus 0.4em\relax New Jersey: Prentice-Hall, Inc., 1975.

\end{thebibliography}

\end{document}